\documentclass{aa}  
\usepackage{graphicx}
\usepackage{epstopdf} 
\usepackage[varg]{txfonts}
\usepackage{lscape}
\usepackage{natbib}
\usepackage{rotating}
\usepackage{amsmath}
\usepackage{longtable}
\usepackage{multirow}
\usepackage[]{textpos}   
\usepackage{xcolor}

\def\ie{i.e.\,}
\def\eg{e.g.\,}

\def\Ms{M_{\star}}

\def\Msun{M_{\odot}}

\def\vrot{V_{\mathrm{rot,max}}}
\def\vcirc{V_{\mathrm{circ}}}
\def\disp{\sigma_0}
\def\deg{\ifmmode^\circ\else$^\circ$\fi}
\def\alphaTF{\ifmmode{\alpha_{\mathrm{\,{\small TF}}}}\else{$\alpha_{\mathrm{\,{\small TF}}}$}\fi}

\begin{document}

\title{Formation of S0 galaxies through mergers}
\subtitle{Evolution in the Tully-Fisher relation since $z\sim1$}

\author{Trinidad Tapia\inst{1}, M.~Carmen Eliche-Moral\inst{2}, H\'{e}ctor Aceves\inst{1}, Cristina Rodr\'{\i}guez-P\'{e}rez\inst{4}, \\ Alejandro Borlaff\inst{2,3}  \and Miguel Querejeta\inst{5} }

\institute{Instituto de Astronom\'{\i}a, Universidad Nacional Aut\'onoma de M\'exico, Apdo.~106, Ensenada BC 22800, Mexico, \email{trinidadtapiap@gmail.com}
\and
Instituto de Astrof\'{\i}sica de Canarias, C/ V\'{\i}a L\'actea, E-38200 La Laguna, Tenerife, Spain.
\and
Departamento de Astrof\'{\i}sica, Universidad de La Laguna, E-38200 La Laguna, Tenerife, Spain.
\and
Departamento de Astrof\'{\i}sica y CC.~ de la Atm\'osfera, Universidad Complutense de Madrid, E-28040 Madrid, Spain.
\and
Max-Planck-Institut f\"{u}r Astronomie, K\"{o}nigstuhl, 17, 69117 Heidelberg, Germany.
}
\date{} 
\abstract{
Lenticular (S0) galaxies are known to derive from spiral galaxies. The fact that S0s nearly obey the Tully-Fisher relation (TFR) at $z\sim 0$ (as spirals have done in the last $\sim$9\,Gyr) is considered an argument against their major-merger origin because equal mergers of two disc galaxies produce remnants that are outliers of the TFR.
}
{
We explore whether a scenario that combines an origin by mergers at $z\sim 1.8-1.5$ with a subsequent passive evolution of the resulting S0 remnants since $z \sim 0.8$--1 is compatible with observational data of S0s in the TFR both at $z \sim 0.8$ and $z \sim 0$.
}
{
We  studied a set of major and minor merger experiments from the GalMer database that generate massive S0 remnants that are dynamically relaxed and have realistic properties. We  analysed the location of these remnants in the photometric and stellar TFRs assuming that they correspond to $z \sim 0.8$ galaxies. We  then estimated their evolution in these planes over the last $\sim 7$\,Gyr considering that they have evolved passively in isolation. The results were compared with data of real S0s and spirals at different redshifts. We  also tested how the use of $\vcirc$ or $\vrot$ affects the results.
}
{
Just after $\sim1$--2\,Gyr of coalescence, major mergers generate S0 remnants that are outliers of the local photometric and stellar TFRs (as already stated in previous studies), in good agreement with observations at $z \sim 0.8$. After $\sim 4$\,--7\,Gyr of passive evolution in isolation, the S0 remnants move towards the local TFR, although the initial scatter among them persists. This scatter is sensitive to the indicator used for the rotation velocity: $\vcirc$ values yield a lower scatter than when $\vrot$ values are considered instead. In the planes involving $\vrot$, a clear segregation of the S0 remnants in terms of the spin-orbit coupling of the model is observed, in which the remnants of retrograde encounters overlap with local S0s hosting counter-rotating discs. The location of the S0 remnants at $z\sim 0$ agrees well with the observed distribution of local S0 galaxies in the $\disp$--$M_K$, $\vcirc$--$\disp$, and $\vrot$--$\disp$ planes.
}
{
Massive S0 galaxies may have been formed through major mergers that occurred at high redshift  and have later evolved towards the local TFR through passive evolution in relative isolation, a mechanism that would also contribute to the scatter observed in this relation.
}

\titlerunning{Evolution of S0s in the Tully-Fisher relation}    
\authorrunning{Tapia et al.}

\maketitle

\section{Introduction}\label{secc:introduction}

It has been found that lenticular galaxies (S0s) constitute a very heterogeneous population with a wide range of physical properties \citep[see][]{Laurikainen2010,Cappellari2011,Kormendy2012}. The question of what physical processes are required to form these galaxies in now being  hotly debated. Lenticular galaxies are known to come from the transformation of spiral galaxies \citep[see e.g.~][]{Laurikainen2010}. Cluster environmental processes such as ram-pressure striping, strangulation, and harassment \citep[\eg][]{Gunn1972,Larson1980,Moore1996,Quilis2000,Peng2015} are often considered the main drivers of this transformation. However, numerical studies have shown that mergers, in particular major ones (mass ratios from 1:1 to 3:1), can also give rise to remnant galaxies that can be classified as S0s in terms of their morphology, structure, and scaling relations \citep[][]{Bekki2001a,Bekki2011,Eliche-Moral2011,Eliche-Moral2012,Eliche-Moral2013, Borlaff2014,Tapia2014,Querejeta2015,Querejeta2015b}. Furthermore, \citet{Athanassoula2016} have recently shown that major mergers can build up remnants of even later types (see also \citealt{Springel2005c}). 

Considering that S0s are descendants of spiral galaxies, the effects of environmental processes on the velocity field of the spiral precursor (in particular those of ram-pressure stripping) are expected to be relatively low and difficult to observe \citep{Kronberger2008a}, whereas mergers would noticeably affect the kinematics and mass distribution of the progenitor galaxies \citep[even minor ones, \ie with mass ratios $>$ 7:1, see][]{Hernquist1992b,Eliche-Moral2006,Kronberger2006,Sarzi2015}. Thus, a possible way to distinguish among these formation processes is by analysing the kinematics of spirals and S0s. 
 
A good indicator of the global kinematics in galaxies is the Tully-Fisher relation (TFR), which relates the intrinsic luminosity and the rotational velocity of a galaxy \citep{Tully1977}. It has been widely used to study the properties of spirals \citep[e.g.][]{Kauffmann1999,Aceves2005,Cresci2009,Torrey2014} and, to a lesser extent, of S0s \cite[\eg][]{Bedregal2006,Williams2010a,Davis2016}.

At low redshift,  spirals and S0s both follow a similar TFR, but whether there is actually an offset or a different degree of scatter between them is still under debate \citep[see][]{Neistein1999,Bedregal2006,denHeijer2015}. Some authors consider that an offset would be evidence that S0s derive from spirals through simple fading after the gas is stripped through environmental processes \citep[\eg][]{Mathieu2002}. However, simple truncation of star formation in spirals would not be enough to explain the offset \citep{Bedregal2006,Noordermeer2008} or the higher dispersion of the S0s in the TFR \citep{Hinz2001}. Even in cluster environments, other authors
suggest that  more complex formation mechanisms such as minor mergers, tidal interactions, or slow encounters could explain these differences \citep[\eg][]{Hinz2003}. 

At high redshift, spirals also follow the TFR (at least until $z\sim 1$), but offset from the local TFR by $\sim 0.4$~mag \citep{Boehm2015}. For S0s, the only study available so far is \citet{Jaffe2014}, who have found that bright elliptical and S0 galaxies with measured emission lines at $z < 1$ follow a $B$-band TFR that is 1.7~mag fainter than the TFR of their spiral counterparts. Although it is unclear whether all S0s at $z \sim 1$ really obey a TFR offset from the local one or if they are just strong outliers of it, it is obvious that S0s are scattered from the TFR of spirals both at high and low redshifts and that they have evolved towards the local TFR in the last $\sim 7$\,Gyr. What mechanisms could explain the observed distribution of the S0 population in the Tully-Fisher (TF) plane both at high and low redshifts and their migration in this plane?

To address this question, we tested the merger scenario, especially major encounters. Numerical studies have found that minor mergers only produce a mild decrease in the angular momentum of stellar discs, contributing to the scatter observed in the TFR of S0s, but they are incapable of creating a strong outlier unless the galaxy has accreted several satellites \citep{Qu2010}. Conversely, major mergers can transform spiral galaxies initially obeying the TFR into a remnant typically in an outlier location \cite[\eg][]{Kauffmann1999,Kronberger2007,Covington2010,DeRossi2012}. \citet{Covington2010} analysed the location of intermediate and advanced stages of major and minor mergers in the stellar TFR, an analogue of the traditional TFR in which the luminosity is replaced by the stellar mass of the galaxy ($\log(\vrot)$ -- $M_\star$). Their experiments were evolved typically for $\sim 4$\,Gyr and the authors located the final remnants at different redshifts to simulate the observational biases. They found that ongoing mergers and early remnants of major encounters are \emph{always outliers of the stellar TFR}, independent of the redshift at which the remnants were observed. Unfortunately, these authors did not investigate the location of the remnants in the photometric TFR or considered their true morphological type, nor did they analyse their evolution after the merger.

In addition, many observational studies point to a global fading of spirals and S0s for a given rotational velocity as time goes by, usually attributed to the simple fading of their stellar populations \citep{Simard1998,Neistein1999,FernandezLorenzo2009}. However, no study has explored a formation scenario for S0s that combines major merger simulations with a realistic treatment of the passive evolution undergone by the complex stellar populations in the remnant and the effects of the residual star formation in them. Therefore, we explore this scenario by analysing a set of N-body major and minor merger simulations that produce realistic S0 remnants, taking into account how their stellar populations would have evolved through passive evolution in isolation until the present. 

This paper is organized as follows. In Sect.~\ref{secc:simulations} we briefly describe the simulations and in Sect.~\ref{secc:methodology} we explain our analysis procedure. In Sect.~\ref{secc:results} we analyse the location of the remnants in the photometric and stellar TFRs at $z\sim 0.8$ and how it changes through passive evolution down to $z\sim 0$. We also compare the location of our remnants at $z\sim 0$ with real S0s in the central velocity dispersion ($\disp$) vs.~total $K$-band absolute magnitude ($M_K$) plane, the circular velocity ($\vcirc$) vs.~$\disp$ plane, and the rotational velocity ($\vrot$) vs.~$\disp$ plane. In Sect.~\ref{secc:discussion} we discuss our results in the framework of current observational studies regarding the TFR of early-type galaxies. Finally, in Sect.~\ref{secc:conclusions} we summarize our main results and give conclusions. 

\section{Description of the simulations}
\label{secc:simulations}

Our sample of merger experiments was extracted from the GalMer database\footnote{GalMer is publicly available at: \url{http://www.project-horizon.fr/}}, which is a public library of N-body simulations of galaxy mergers, developed at intermediate resolution with a wide range of morphological types, mass ratios, and orbital parameters. We refer the reader to \citet{Chilingarian2010} for a full description, but here we provide a brief description.

The progenitor galaxies correspond to five different morphological types (E0, S0, Sa, Sb, and Sd) with two mass ranges (giants and dwarfs -- g, d). The dwarfs are ten times less massive than the giants. The mass ratio of the merging experiments spans from 1:1 to 3:1 for the major mergers and between 7:1 to 20:1 for the minor ones. Different initial velocities and pericentre distances, disc inclinations with respect to the orbital plane (six possible angles), and two spin-orbit orientations (prograde and retrograde) are considered. Each experiment has a duration of $\sim3$ to 3.5\,Gyr depending on the model.

The total stellar mass in the giant progenitors varies from $\sim 0.5$\,--\,$\mathrm{1.5 \times 10^{11}\,\Msun}$ and the gas mass goes from $\mathrm{9.2 \times 10^{9}\,\Msun}$ to $\mathrm{1.7 \times 10^{10}\,\Msun}$, the gSd being the most gas-rich progenitor. The gE0 and gS0 progenitors are devoid of gas, as are their dwarf analogues.

The giant galaxies have a total of 120\,000 particles distributed among collisionless stellar particles (we  refer to them as old stellar particles), hybrid particles (we refer to their gaseous mass content as gas and to their stellar mass content as young stellar particles), and dark matter particles. The dwarf galaxies have 48\,000 particles.

The galaxy-galaxy evolution was performed using a Tree-SPH code \citep{Semelin2002}. The adopted softening length is $\mathrm{\epsilon = 280\,pc}$ for all the components in the giant-giant interactions and $\mathrm{\epsilon = 200\,pc}$ for the giant-dwarf ones.

The star formation efficiency for the SPH particles was parametrized using a hybrid method \citep{Mihos1994}. The effects of star formation on the interstellar medium such as stellar mass loss, metallicity enrichment of the gas, and the energy injection by supernova explosions were also taken into account using the technique described in \citet{Mihos1994}. 

The old stellar component of the giant galaxies has a radial metallicity profile \citep{Kennicutt2003,Magrini2007,Lemasle2008}. This is also the case for the dwarf galaxies, but the dependency with their mass and half-mass radius is taken into account \citep{Tremonti2004,Lee2006}. This metallicity remains constant during the simulations, whereas the metallicity of both the gas and the new stellar component changes with time, due to the metal enrichment of the gas by supernovae explosions.
  
We denote each merger experiment using the following convention: g[\textit{type$_{1}$}]g[\textit{type$_{2}$}]o[\textit{orbit}], where \textit{type$_{1}$} and \textit{type$_{2}$}  represent the morphological type of each of the merging galaxies and \textit{orbit} corresponds to the tag number for the orbit used in the GalMer database.

\section{Methodology}
\label{secc:methodology}

We identified and selected those merger experiments from the GalMer database in which the remnant system at the end of the simulation was dynamically relaxed and had a morphology, star formation level, and gas content typical of real S0 galaxies. Specific details on the criteria used to select the sample will be provided elsewhere (Eliche-Moral et al., in prep.), but \citet{Borlaff2014} and \citet{Querejeta2015} contain a summary of how this process is done. Hence, in Sect.\,\ref{secc:sample} we only mention the aspects that are relevant to the present study. 

We obtained the absolute magnitudes in the $B$ and $K$ bands of the remnants considering the light emitted by the old stellar particles and the young stellar particles born during the interaction, assuming that the final remnant is observed at $z=0.8$ and computing the effects of passive evolution since that epoch. The effects of the residual star formation and gas remaining at the final moment of the simulation were also accounted for (Sect.\,\ref{secc:magnitudes}). Then, we derived their global kinematical parameters simulating long-slit spectroscopic data to obtain the maximum rotational velocity ($\vrot$), the circular velocity ($\vcirc$), and the central velocity dispersion ($\disp$) of each remnant (Sect.\, \ref{secc:kinematics}). We thus compared the location of our S0 remnants in the photometric and stellar TF diagrams with real S0 and spiral galaxies at $z=0.8$ and $z=0$ (Sect.\,\ref{secc:results}).

\subsection{Selection of S0s in artificial photometric images}
\label{secc:sample}

\begin{figure*}[t!h]
\centering
   \includegraphics[width=18.3cm]{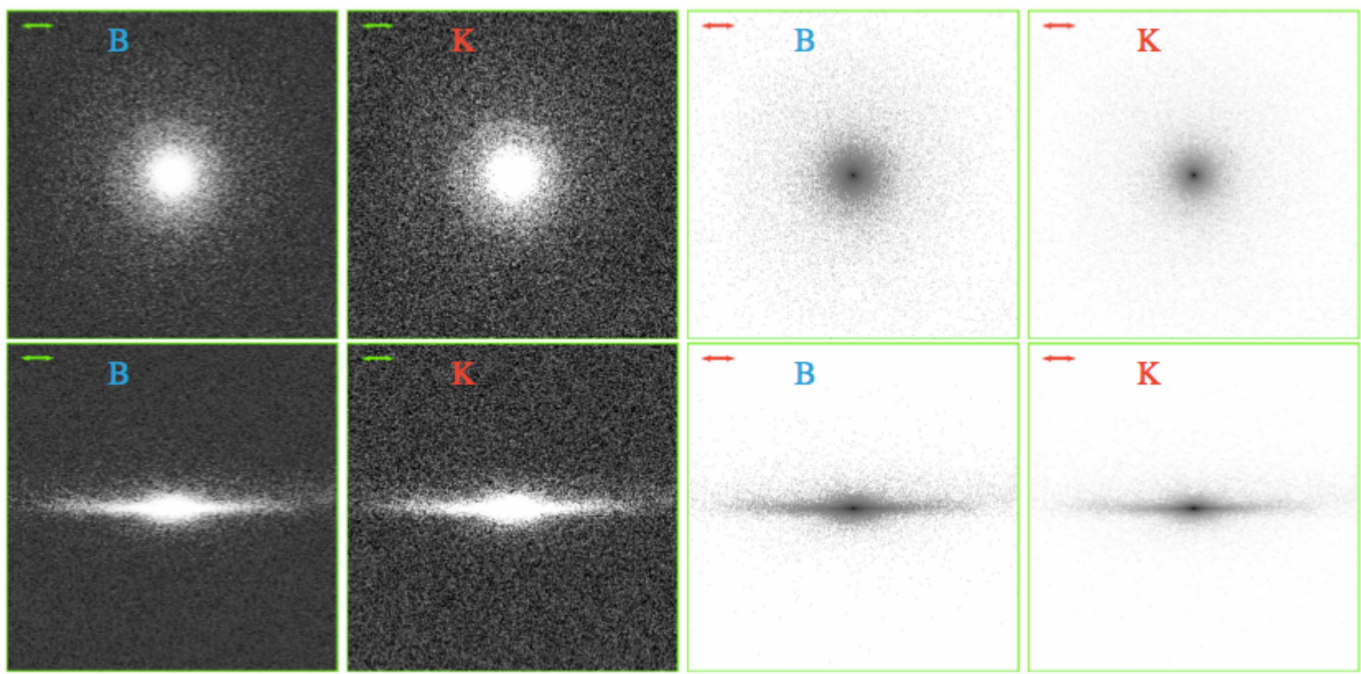}
  \caption{Simulated photometric images in the $B$ and $K$ for the remnant of experiment gSagSao1, assuming a distance of 30\,Mpc. \emph{First and second columns}: Face-on and edge-on views of the remnant using a logarithmic grey scale to highlight the structure in the outskirts of the galaxy. \emph{Third and fourth columns}:  Same as in the previous rows, but using an inverted logarithmic grey scale to enhance the substructures in the core of the remnant. The spatial resolution is 0.7\arcsec. The field of view is $\mathrm{50 \times 50\,kpc}$. The segment in the top left corner of each panel represents a physical length of 5~kpc.}
\label{fig:gSagSao1}
\end{figure*} 

To ensure that our merger remnants would be considered as an S0 galaxy by an observer, we simulated realistic photometric images in the bands $B$ and $K$, mimicking the general properties of current observational surveys of nearby S0 galaxies (seeing, signal-to-noise ratio, spatial resolution, and magnitude limit), both for face-on and edge-on views of the remnant. For the $K$-band images we simulated the conditions of the Near-InfraRed S0 Survey (NIRS0S,\,\citealt{Laurikainen2011}) to perform the comparisons shown in \citet{Querejeta2015}. We added photon noise to the model images to have a signal-to-noise ratio ($S/N$) of 3 for the limiting magnitude of the images. We placed the model images at a distance of $\sim 30$\,Mpc, which is the average of the NIRS0S survey. We adopted a seeing with a full width at half maximum (FWHM) of  0.7\arcsec\, and a limiting magnitude of $\mathrm{22\,\,mag\,arcsec^{-2}}$. According to this survey, this is equivalent to an intrinsic spatial resolution of $\sim100$\,pc. We did the same for the images in the Johnson-Cousin $B$ band, mimicking the average observational conditions of the Sloan Digital Sky Survey (SDSS, \citealt{Abazajian2003}). We considered a $S/N=5$ for the limiting magnitude of 26\,mag\,arcsec$^{-2}$. The seeing was also set to a FWHM of 0.7\arcsec. In the images, we have transformed the intrinsic physical length into angular distance and we have accounted for the effects of cosmological dimming using a $\Lambda$CDM concordant cosmology ($\Omega_{M} = 0.3$, $\Omega_{\Lambda} = 0.7$, $\mathrm{H_{0} = 70\,km\,s^{-1}\,Mpc^{-1}}$).  

Five co-authors independently inspected these images and assigned a morphological type according to their visual morphology in these images: E, E/S0, S0, or S (elliptical, elliptical/lenticular, lenticular, or spiral, respectively). The final morphological type assigned to each remnant is the median of the five classifications. In $\sim85\%$ of the cases the authors all made the same classification. Then, the final sample of S0-like remnants used in this study consists of 71 merger remnants, 42 corresponding to major encounters and 29 to minor ones. Among the total, $69\%$ correspond to remnants of direct encounters and 31\% from retrograde encounters. These 71 S0-like remnants are analogous to real S0s in terms of morphology, structure, star formation rates (SFRs), and gas content (Eliche-Moral et al., in prep). We note that all our minor mergers are onto the same gS0 progenitor. Additionally, our sample is smaller than the one presented in \citet{Borlaff2014} and \citet{Querejeta2015} because we excluded the merger remnants classified as E/S0 to focus on the well-defined S0s.

In Fig.\,\ref{fig:gSagSao1} we show the simulated photometric images in the $B$ and $K$ bands for one of the remnants classified as S0 in face-on and edge-on views. The effects of dust extinction were not simulated in these images because they were developed to perform only a morphological classification, but we did consider them to study the TF planes (see Sect.\,\ref{secc:magnitudes}).

\citet{Qu2010} have already analysed the effects of minor mergers on the TFR using the models of the GalMer database. Although the present study focuses on the effects of major merger events in combination with a later passive evolution of the system, we also included the minor mergers in the analysis for completeness. On the other hand, we assumed that mergers ended at $z\sim 0.8$ and that the remnants have evolved in isolation since then, whereas Qu et al. locate the final remnants of the minor mergers at $z=0$ directly. Thus, our procedure deals with a different formation and evolution scenario.

We also checked that the disc-like progenitors (gS0, gSa, gSb, gSd) approximately fulfil the stellar TFR of spiral galaxies to guarantee that we are coherent with the scenario to be tested. To show this, we have overplotted the location of these progenitors in the stellar TFR (Fig.\,\ref{fig:vcircmass}) in Sect.\,\ref{secc:results}.

\subsection{Time evolution of the absolute magnitudes of the S0 remnants}\label{secc:magnitudes}
\begin{figure}[t!h]
\centering
   \includegraphics[width=8.9cm]{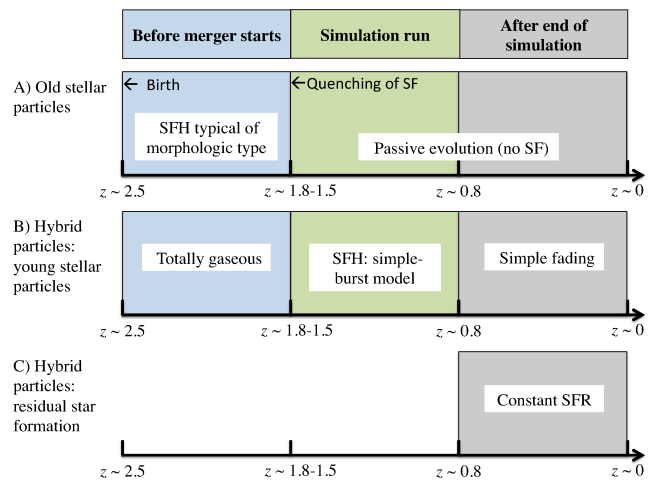}
  \caption{Schematic timeline of the star formation of each type of particle that contributes to the luminosity of the galaxy model since  $z \sim 2.5$. We divided this according to the relevant stages of our  scenario: before the merger simulation, the simulation run, and the subsequent evolution. We indicate the approximated redshift at which each of these stages occurs.}
\label{fig:schematicSFH}
\end{figure}

To calculate the photometric magnitudes in the Vega system, the mass of the particles has been converted into light flux considering the mass-to-light ratio ($M/L$) in each band.  In Fig. \ref{fig:schematicSFH} we present a schematic timeline that summarizes how we have accounted for the star formation (SF), the ageing of the stellar populations and the effects of the residual star formation in the remnants, which all together determine the total luminosity of the galaxy models.  The timeline of each type of particle, \ie the old stellar particles (A) and the hybrid particles (B and C) are shown separately. The relevant temporal stages of the scenario we are testing are shown in different colours.

Based on the average age estimated for the old stellar populations in the discs of present-day S0s \cite[see][]{Silchenko2012,Silchenko2013}, we consider that all the old stellar particles have an age of 11\,Gyr at $z=0$ regardless of the progenitor they come from. This assumption thus sets the age of the old stellar particles at any time, and in particular at $z\sim 0.8$, when we consider the remnant to be already in place. It also establishes the epoch at which these stars were born ($z\sim 2.5$) and when the merger had to start ($z\sim 1.8-1.5$), since the merger experiments have a fixed duration (typically $\sim 3-3.5$\,Gyr).

We consider that the old stellar particles in each progenitor galaxy have experienced a star formation history (SFH) proportional to their mass and determined by the morphological type of their progenitor (E, S0, Sa, Sb or Sd). 
The SFH assumed for each morphological type was chosen according to observations \citep[see][and references therein]{Eliche-Moral2010}.  We used the stellar population synthesis model by \citet{Bruzual2003}, considering a Chabrier initial mass function \citep{Chabrier2003}, and the evolutionary tracks of the Padova Group \citep{Bertelli1994}.
Each old stellar particle evolves from the moment its stellar content is assumed to have been born (at $z\sim 2.5$), until it suddenly quenches when the merger simulation starts ($z\sim 1.8-1.5$). 
From this moment until the present these particles are assumed to evolve passively (see Fig.~\ref{fig:schematicSFH}), and star formation is transferred to the hybrid particles during the merger.  
 This complex SFH determines the value of the M/L of the old stellar particle at each redshift.

The SFH of the hybrid particles, which are totally gaseous at the beginning, is different for each one and evolves with time in a complex manner during the simulation (see Fig. \ref{fig:schematicSFH}). We approximated the SFH of each hybrid particle by a single-burst stellar population model with an age and metallicity equal to the mass-weighted average value that the particle has at the end of the simulation. We estimated the $M/L$ in each band and converted the stellar mass content in the particle at the end of the simulation into light flux in that band using the \citet{Bruzual2003} models.

To account for the effects of the ageing of the stellar populations in the merger remnants  for the last $\sim 7$\,Gyr (from $z\sim 0.8$ to $z\sim 0$), we estimated the changes in luminosity of our final S0-like remnants assuming that they correspond to $z\sim 0.8$ galaxies and that their stellar populations have evolved similarly to a `closed box' model. We used the \citet{Bruzual2003} models to derive the evolution of the $M/L$ in these bands for each old stellar particle in the remnant assuming they keep evolving passively. 

As noted before, we approximated the complex SFH of the young stellar particles formed during the simulation by simple stellar populations with the average stellar age and metallicity of the particle at the end of the simulation ($z\sim 0.8$). The contribution to the total luminosity at any time since $z\sim 0.8$ down to the present was determined following the simple fading of these models (see Fig. \ref{fig:schematicSFH}).

Many of our S0-like remnants have residual star formation in their centres at the final moment of the simulation due to starbursts triggered by gas inflows produced by the interaction \citep{Borlaff2014,Querejeta2015,Querejeta2015b}. To account for the effects of this residual star formation on the magnitudes of our remnants, we assumed that this SFR remains constant with time and ends when the gas reservoir is finished (see Fig. \ref{fig:schematicSFH}, panel C). We also used the stellar population synthesis models by \citet{Bruzual2003} to estimate the contribution in each band of the light coming from this residual star formation. In this case, the input for these models is the average metallicity of the gas content available in the hybrid particles at the end of the simulation. We  derived the evolution of the $M/L$ in the $B$ and $K$ bands of the stellar population emerged from this relic SFR and included this contribution to the total luminosity of the remnant at each time.

\begin{figure}[t!h]
\centering
   \includegraphics[width=8.8cm, bb = 10 220 270 420, clip]{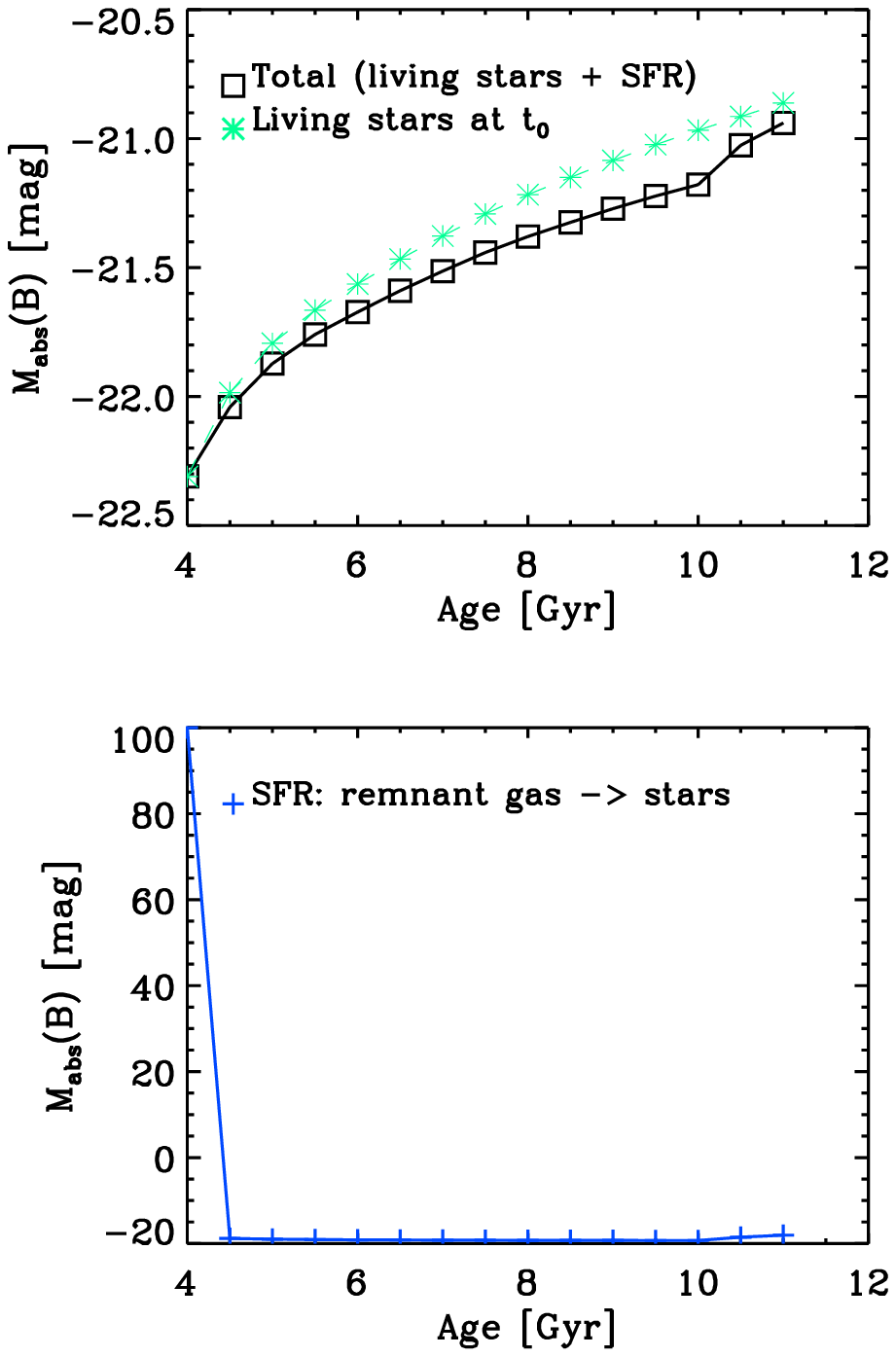}
  \caption{$B$-band luminosity evolution of the S0 remnant labelled gSbgSdo5 for a time period of 7\,Gyr, which corresponds to the phase of passive evolution in isolation of the remnants in the scenario we are testing. The evolution of the absolute $B$ magnitude is plotted as a function of the age of the old stellar population. We assume that the old stellar particles are $\sim 11$\,Gyr old at present, in agreement with data of local S0s (see references in the text), so they have an age of 4\,Gyr at $z\sim 0.8$. Black squares:  time evolution of the total $B$-band magnitude of the S0 galaxy, accounting for the old stellar particles, the young stellar particles, and the residual star formation in the remnant. Green asterisks: contribution to the total luminosity at each epoch of the light coming from the stellar content already existing at $z\sim 0.8$ (living stars, i.e. old$+$young stars).} 
\label{fig:passiveevolution}
\end{figure} 

In Fig.\,\ref{fig:passiveevolution} we show the evolution of the total $B$-band absolute magnitude for an S0-like remnant of an encounter with an initial gas fraction of 0.23, from the final moment of the simulation ($z\sim 0.8$) down to the present. Instead of using the cosmic time, we represent the magnitudes as a function of the age of the old stellar content (assumed to be equal to 4\,Gyr at $z=0.8$, thus $\sim 11$\,Gyr at $z=0$). We plot the magnitudes that correspond only to the evolution of the stellar content already in place at $z\sim 0.8$ (the `living stars', \ie the old and young stellar particles), as well as the total $B$ magnitude accounting for the effects of the residual star formation. In this case, the contribution of the relic star formation to the total flux of the galaxy can amount up to $\sim 20$\% of the luminosity in the $B$ band $\sim 5$\,Gyr after $z\sim 0.8$ (because the stars are very young). Nevertheless, at that moment, the gas remaining in the remnant is totally consumed and the residual SFR quenches. From that moment, the contribution of the stars formed through this relic SFR fades in the $B$ band. This is why the magnitudes in which we considered this relic SFR tend to converge to the ones in which this effect is not taken into account for the last 2\,Gyr of the computed evolution.

\subsection{Corrections to the total absolute magnitudes}
\label{secc:corrections}

\subsubsection{Dust extinction}
\label{secc:dust}

We applied a dust extinction correction to our magnitudes considering the average dust extinctions for observed local S0s in the $B$ and $K$ bands, namely $A_B\sim 1.2$\,mag and $A_K\sim 0.15$\,mag, assuming that all our S0-like remnants have the same extinction at all redshifts (see the justification in  Appendix \ref{appendix:dust}). In Table\,\ref{tab:kinematicsz0} we list the final $B$- and $K$-band absolute magnitudes of our S0-like models including the effects of dust extinction for $z \sim 0.8$ and $z \sim 0$. 

\subsubsection{k-correction}
\label{secc:kcorrection}

Observational data at intermediate to high redshifts also require a k-correction for getting rest-frame absolute magnitudes in a given photometric band, even those studies that use the observed band that best matches the rest-frame band of interest at the redshift of the galaxy \citep[\eg][]{Ziegler2002,FernandezLorenzo2009}. Our models do not require any k-corrections because we directly compute the magnitudes in rest-frame spectra. Owing to the uncertainties implicit to the assumptions made on the stellar populations, SFH modelling, the treatment of the residual star formation, and the dust extinction in the S0-like remnants, no errors have been derived for their $B$- and $K$-band absolute magnitudes at $z\sim 0.8$ and $z \sim 0$ in Table\,\ref{tab:kinematicsz0}. Observational photometric errors in broad bands are typically below 0.1\,mag, but these errors are clearly underestimated as the uncertainties in the dust- and k-corrections of real S0s are much higher. Consequently, we assume a typical uncertainty of $\sim 0.2$\,mag in our remnants to give room to the comparison of the models with real data.

\subsection{Kinematics of the remnants}
\label{secc:kinematics}

Although we have estimated the time evolution of the luminosity of the remnants for a period of 7 Gyr after the last snapshot available for each model in GalMer (which we assume to correspond to the state of the remnant at $z\sim 0.8$), we assume that the kinematical quantities like $\vrot$, $\vcirc$, and $\disp$ barely change once the remnant is already dynamically relaxed by that epoch to avoid the artificial numerical heating inherent to evolving these models for 7\,Gyr more. In major mergers, dynamical relaxation happens just $\sim 0.5$\,Gyr after the full merger and from this moment, the kinematical quantities do not change appreciably \citep[see also ][]{DiMatteo2008b,Covington2010}. Considering that this is a much shorter time period than the total time experienced by our remnants after the full merger ($\sim 1$--2.5\,Gyr), our systems must be in equilibrium. However, we have specifically selected only the S0-like remnants that were fully merged and dynamically relaxed according to quantitative criteria, ensuring that our S0-like remnants are in total dynamical equilibrium (Eliche-Moral et al., in prep.).

To test this assumption, we evolved a representative sample of the remnants used here 5~Gyr further in time by means of the numerical code \texttt{GADGET} \citep{Springel2005}. In all cases, the energy conservation was better than 0.5$\%$, and the virial ratio ($2T/|U|$) maintained a value $\approx 1$. Moreover, no bar-like structures were developed, only some weak and transient spiral arm features were observed, so the systems were really in dynamical equilibrium. However, while numerical heating affected the $\vrot$ values negligibly (by $\lesssim 15\%$), it artificially increased the $\disp$ values of the evolved galaxies by $\sim 30$--50$\%$. Considering that the $\disp$ and $\vrot$ values of our S0-like remnants change by $\lesssim 5\%$ during the $\sim 1$--2\,Gyr from the full merger until the last time computed for each model (corresponding to the remnant at $z \sim 0.8$), we assume that the kinematical quantities of the remnants at $z\sim 0.8$ must barely change during their evolution in isolation for the next 7\,Gyr, as supported by many studies \citep[see][]{Sellwood2002,Roskar2008a,Roskar2008b,Minchev2013,VeraCiro2014,Sellwood2014}.

\begin{figure}[t!h]
\centering
   \includegraphics[width=8.8cm]{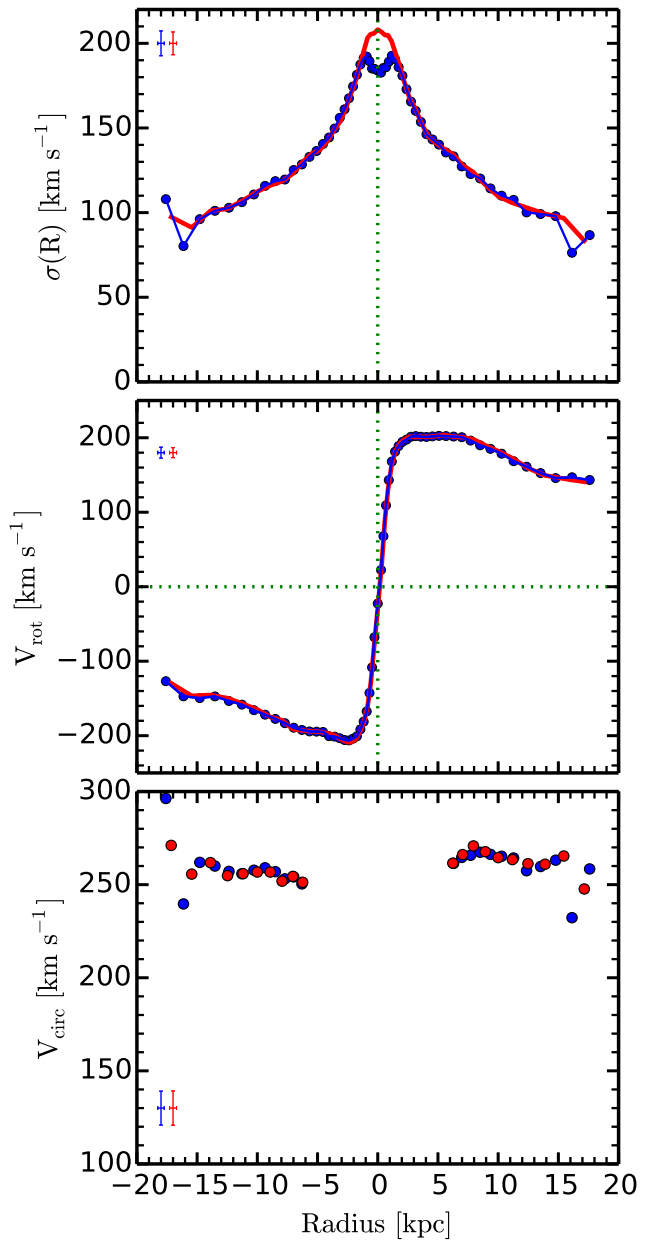}
  \caption{\emph{Top panel}: Velocity dispersion profile of the S0 remnant of the merger experiment gSagSao9. \emph{Middle panel:} Rotation curve for the remnant. \emph{Bottom panel:} Circular velocity measurements (after applying the asymmetric drift correction to its rotation curve) used to derive the circular velocity of this S0 remnant, as explained in Sect.\,\ref{secc:calculovcirc}. In the three panels, the red circles correspond to the values obtained by weighting the data by the stellar mass of the particles, whereas the blue correspond to the luminosity-weighted values. The error bars in each panel corresponds to the width of the radial bin and the statistical error on the y-axis.}
\label{fig:rotcLgSagSao9tf.eps}
\end{figure}

\subsubsection{Estimates of the maximum rotational velocity and central velocity dispersion}
\label{secc:calculovrot}

We computed radial profiles of the rotation velocity and velocity dispersion mimicking real 1D spectroscopic observations for each final S0-like remnant. We placed a virtual slit along the major photometric axis of the remnant disc in an edge-on view (to avoid inclination corrections) and estimated the weighted average and standard deviation of the line-of-sight velocity distribution (LOSVD) of the particles projected into spatial bins of 1\,kpc along the slit. We considered all stellar material within a radial extent of 20\,kpc from the remnant's centre. The length and width of the slit is enough to enclose the disc in all our S0-like remnants. The velocity of each particle with stellar content (collisionless or hybrid) is weighted by its stellar mass or by the light it emits (light-weighted estimates) to derive the line-of-sight average rotation velocity and dispersion in each bin. Therefore, we obtained mass- and light-weighted rotation curves and velocity dispersion profiles. The uncertainty in these profiles was estimated considering the statistical errors on the particle count in each bin \citep{Gehrels1986}.  

In Fig.~\ref{fig:rotcLgSagSao9tf.eps} we show the rotation curve ($V_\mathrm{rot}(R)$), the velocity dispersion ($\sigma(R)$) profile, and the circular velocity ($V_\mathrm{circ}(R)$) after the asymmetric drift correction (see Sect.\,\ref{secc:calculovcirc}) of one S0-like remnant. We represent the light- and mass-weighted profiles. In general, these profiles are similar to those observed in real S0 galaxies, independently of how the numerical data are weighted. In fact, the way the data are weighted barely modifies these curves except in the central region where the new stellar particles have accumulated at the end of the merger. The luminosity-weighted values of the velocity dispersion in the central region are lower than the mass-weighted values because of the existence of an over-luminous co-rotating substructure formed in the centre of this remnant in particular.

Many other remnants also show peculiar features in the central parts of their rotation curves and velocity dispersions profiles, due to the presence of kinematically-decoupled inner components, such as inner co- and counter-rotating discs and kinematically-decoupled cores (KDC). This was expected: several studies have pointed out that major mergers are the most likely mechanism for generating KDC galaxies \cite[e.g.][]{Balcells1999,McDermid2006,Tsatsi2015} and minor mergers can produce dynamically cold inner components such as co- and counter-rotating inner discs, bars, rings, or pseudorings \citep{Eliche-Moral2011,Mapelli2015}. 

We estimated the maximum value of the rotation curve, $\vrot$, and the central velocity dispersion, $\disp$, for each S0-like remnant from both the mass- and light-weighted profiles. These values and their errors are shown in Table\,\ref{tab:kinematicsz0} for the S0 remnants and their giant progenitor galaxies.  In Appendix~\ref{appendix:weight} we comment in detail how the weighting of the data affects the results, and we justify our choice for a luminosity-weighted kinematics (see the results in Sect.\,\ref{secc:results}). 

\subsubsection{Estimates of the circular velocity}
\label{secc:calculovcirc}

We also computed the circular velocity ($\vcirc$) for each S0-like remnant (both for luminosity- and mass-weighted data) since this quantity is very often used to study the TFR (further details in Sect.~\ref{secc:remarks}). To determine $\vcirc$ we followed the procedure described in \cite{Neistein1999}, who applied two corrections to the rotation curves, the first to correct for the integration along the line-of-sight, and the second to account for the non-circular orbits of the stars. We neglected the first correction because our models are already in an edge-on position, but we applied the asymmetric drift correction. The circular velocity ($V_{circ}(R)$) is given by

\begin{equation}\label{ecc:vcircN99}
V_{\mathrm{circ}}(R)^{2} = V_\mathrm{\phi}(R)^{2} + \sigma_{\phi}(R)^{2}\left[2\left(\frac{R}{R_{\mathrm{disc}}} -  \frac{\partial \ln \sigma^{2}_{R}}{\partial \ln R} \right ) - 1 \right ] ,
\end{equation}

\noindent where $R_{\mathrm{disc}}$ corresponds to the disc scale length of each remnant (see Table 3 in \citealt{Querejeta2015}). In our case, the kinematics in the azimuthal direction ($V_\mathrm{\phi}(R)$ and $\sigma_{\phi}(R)$) correspond to our measurements of $V_{\mathrm{rot}}(\mathrm{R})$ and $\sigma(\mathrm{R})$ (Sect.\,\ref{secc:calculovrot}). Typically, the term $\partial \ln \sigma^{2}_{R}/\partial \ln R$ is small and can be neglected. We thus estimated the $V_{\mathrm{circ}}(\mathrm{R})$ through the following expression:

\begin{equation}\label{ecc:vcirc}
V_{\mathrm{circ}}(R)^{2} = V_\mathrm{rot}(R)^{2} + \sigma (R)^{2}\left(2\frac{R}{R_{\mathrm{disc}}} - 1 \right).
\end{equation}

The final circular velocity is estimated by taking the mean value of all the $\mathrm{V_{\mathrm{circ}}(R)}$ data points  that are located in the flat part of the rotation curve for which $\mathrm{V_{\mathrm{rot}}(R)/\sigma (R) \ge 2.5}$, generally at $\mathrm{1.2 < R/R_{\mathrm{disc}} < 3}$. Some of our S0-like remnants have relatively large $\sigma (R)$, even in their outer regions, and thus have $V_{\mathrm{rot}}(R)/\sigma (R) < 2.5$. For these models, we relaxed the criterion to $\mathrm{V_{\mathrm{rot}}(R)/\sigma (R) > 1.0}$ to compute their $\mathrm{\vcirc}$, but the correction can produce unrealistically high circular velocities in some cases. Therefore, their $\mathrm{\vcirc}$ measurements can only be considered an upper limit.

Out of 71 S0-like remnants 14 still did not satisfy the relaxed criterion, so we did not attempt an asymmetric drift correction and we excluded them from the main analysis. They correspond to the models with no values of circular velocity in Table\,\ref{tab:kinematicsz0}. Nevertheless, we have included them in the planes that involve $\vrot$ or $\disp$. We note that many of our S0-like remnants which come from mergers in retrograde orbits host counter-rotating components (Eliche-Moral et al., in prep.), and therefore have higher $\sigma (R)$ so the approximation of \citeauthor{Neistein1999} to the asymmetric drift correction is less appropriate, and is why the majority of the excluded models correspond to mergers in retrograde orbits (see Table\,\ref{tab:kinematicsz0}).
 
\begin{table}
\small
\caption{Kinematical data of nearby S0s obtained from the kinematics of their PNe and for S0 galaxies hosting counter-rotating discs.}
\label{tab:S0scounterrotS0s}
\centering
\begin{tabular}{l rr rr c}
\hline\hline
\multicolumn{1}{c}{\multirow{2}{*}{Name}} & \multicolumn{1}{c}{$\mathrm{M}_B$} & \multicolumn{1}{c}{$\mathrm{M}_K$} &   \multicolumn{1}{c}{$\mathrm{\sigma}$} &
 \multicolumn{1}{c}{$V_\mathrm{rot}$} &  \multicolumn{1}{c}{\multirow{2}{*}{Reference}} \\
\multicolumn{1}{c}{}    & \multicolumn{1}{c}{[mag]} & \multicolumn{1}{c}{[mag]} &  \multicolumn{1}{c}{[km\,s$^{-1}$]} &  
\multicolumn{1}{c}{[km\,s$^{-1}$]} & \multicolumn{1}{c}{}\\
\multicolumn{1}{c}{(1)}    & \multicolumn{1}{c}{(2)} & \multicolumn{1}{c}{(3)} & \multicolumn{1}{c}{(4)} & \multicolumn{1}{c}{(5)}  & \multicolumn{1}{c}{(6)} 
 \vspace{0.05cm}\\\hline\vspace{-0.3cm}\\
\multicolumn{6}{c}{Hosting counter-rotating discs}
\vspace{0.05cm}\\\hline\vspace{-0.3cm}\\
IC719   & -18.7 & -22.6 & 215  & 110     &  1 \\
NGC2551 & -20.1 & -23.3 & 90   &  56    & 2 \\
NGC5631 & -20.2 & -23.9 & 125  & 101  & 2 \\     
NGC4550 & -17.2 & -22.4 & 84   & 170   & 3,4 \\
NGC3593 & -18.4 & -21.3 & 135  &  72   & 5 \\
NGC4138 & -18.8 & -22.8 & 130  &  28   & 6 \\
NGC7332 & -19.7 & -23.0 & 140  &  60   & 7 \\
NGC3941 & -19.9 & -26.4 & 140  & 130  & 8 \\
NGC4546 & -19.8 & -23.9 & 270  &  70   & \hspace{0.05cm}9                            
\vspace{0.05cm}\\\hline\vspace{-0.3cm}\\
 \multicolumn{6}{c}{From the kinematics of their PNe}
\vspace{0.05cm}\\\hline\vspace{-0.3cm}\\
NGC1023 & -21.0 & -23.8 &  84.4   &  236.2    & 10 \\           
NGC2768 & -21.3 & -24.5 &  213.7  &  250.1    & 11 \\           
NGC3115 & -19.9 & -24.0 &  272.7  &  361.4    & 12\\            
NGC3489 & -19.3 & -23.0 &  114.2  &  128.1    & 12 \\           
NGC3384 & -19.0 & -23.6 &  129.9  &  161.1    & 12 \\           
NGC6340 & -20.0 & -23.0 &  100.0  &  50.0     & 13\\            
NGC7457 & -19.2 & -22.4 &  75.8   &  108.0    & 14\\\hline      
\end{tabular} 
\tablefoot{Columns: (1) Galaxy name. (2) Absolute magnitude in the $B$ band from HyperLeda, estimated from the apparent $B$ magnitude and the distance modulus corresponding to the redshift of the galaxy. The apparent magnitudes have been corrected for galactic extinction, internal extinction, and k-correction. (3) Absolute magnitude in the $K$ band, estimated as described in the text. (4) Central velocity dispersion. (5) Maximum rotation velocity. (6) Reference for the data.}
\tablebib{
(1)~\citet{Katkov2013}  ; (2) \citet{Silchenko2009}; (3) \citet{Rubin1999}; (4) \citet{Whitmore1985};
(5) \citet{Bertola1996}; (6) \citet{Thakar1997}; (7) \citet{Fisher1994}; (8) \citet{Fisher1994a};
(9) \citet{Bettoni1991}; (10) \citet{Noordermeer2008}; (11) \citet{Forbes2012a}; (12) \cite{Cortesi2013}; (13) \cite{Chilingarian2009}; (14) \citet{Cortesi2013a}.
}
\end{table}

\begin{figure*}[t!h]
\centering
   \includegraphics[width=18.3cm]{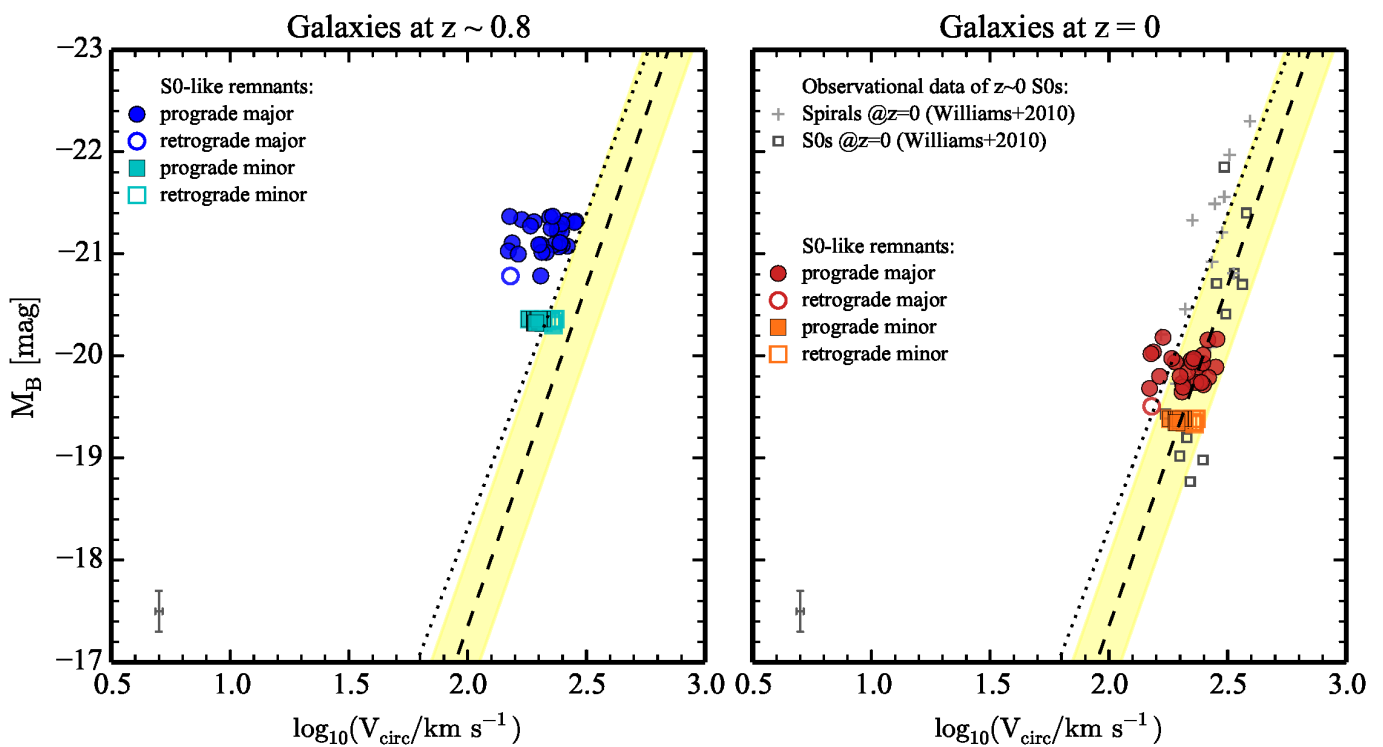}
  \caption{\emph{Left panel:} TFR for our S0-like remnants in the $B$ band at $z\sim 0.8$, using $\mathrm{\vcirc}$. \emph{Right panel:} TFR for our S0-like remnants at $z\sim0$ after $\sim 7$\,Gyr of passive evolution of their stellar populations  (accounting for the effects of the residual star formation in the remnant since $z\sim 0.8$) compared to present-day spiral and S0 galaxies, also using $\mathrm{\vcirc}$. We have distinguished between the remnants of prograde and retrograde encounters. Data are labelled as shown in the figure. In both panels, the dotted line denotes the local TFR in the $B$ band for spiral galaxies by \protect \citet[]{Williams2010a}, whereas the dashed line corresponds to the local TFR in the $B$ band for S0 galaxies also by \protect \citeauthor{Williams2010a}, and the yellow band its $1\sigma$~scatter. The error bar in the bottom left corner of each panel denotes the statistical errors on both axes for our S0-like remnants.}
\label{fig:TFVcircBband}
\end{figure*} 

\subsection{Comparison between observational data and simulations}
\label{secc:remarks}
We compared our models with data of real S0s at different redshifts and in the photometric TFRs considering $B$ and $K$ band magnitudes. In the $B$-band TFR, we used the S0 data at $z\sim 0.8$ by \citet{Jaffe2014} and at $z\sim 0$ by \citet[]{Williams2010a}, both including field and cluster objects; we also used the data of spirals at $z\sim 1$ by \citet{Chiu2007}, the spirals at $0.7<z<1.2$ by \citet{Jaffe2014}, and at $z\sim 0$ by \citet[]{Williams2010a}. In the $K$-band TFR, we compared our models with S0 data at $z\sim 0$ by \citet[field and cluster]{Bedregal2006}, \citet[field and cluster]{Williams2010a}, \citet[cluster]{Rawle2013}, and \citet[from the MASSIVE sample]{Davis2016}. The sample by \citeauthor{Bedregal2006} contains S0 galaxies from the Virgo, Coma, and Fornax clusters; the \citeauthor{Rawle2013} sample is made of early-type galaxies from Coma; and the galaxies in \citeauthor{Davis2016} correspond to gas-rich fast-rotating massive early-type galaxies from the MASSIVE survey. Given that we have not found any trend with environment in the TFR, unless indicated otherwise we represent the data of each galaxy type coming from the same author with the same symbol, whether the galaxy was in the field or in a cluster in most figures. 

We also collected photometric and kinematical data of nearby S0 galaxies that are known to host counter-rotating discs (stellar or gaseous) for both photometric TFRs (in the $K$ and $B$ bands). In Table~\ref{tab:S0scounterrotS0s} we list their data and the corresponding references. Their absolute magnitudes in the $K$ band were derived using the apparent $K$-band magnitudes from the Two Micron All Sky Survey (2MASS) catalogue\footnote{2MASS website at IPAC: http://www.ipac.caltech.edu/2mass/} and distances from the NASA/IPAC Extragalactic Database\footnote{NED website available at: https://ned.ipac.caltech.edu/} (NED). Their k-corrections have been neglected, due to their closeness. Additionally, we selected a sample of local S0 galaxies, whose stellar kinematical data were derived using planetary nebulae (henceforth PNe). We note that the velocity dispersion of these galaxies were not derived directly from the PNe themselves, but using a fitting technique (see the references in Table \ref{tab:S0scounterrotS0s} for more details).  This sample is also included in Table~\ref{tab:S0scounterrotS0s}. Their $K$-band absolute magnitudes were also estimated using data from 2MASS and NED. The absolute $B$-band magnitudes for all galaxies in this table (hosting counter-rotating discs or using kinematical data from their PNe) were obtained from HyperLeda database\footnote{HyperLeda Database webpage: http://leda.univ-lyon1.fr/} \citep{Makarov2014}, and were derived from their distance modulus and apparent $B$-band magnitudes corrected for Galactic extinction, internal extinction, and k-correction. 

A difficulty arising from the comparison of kinematical data of different galaxy samples are the uncertainties entailed by the different techniques used to estimate $\vrot$, $\vcirc$, or $\disp$. The observational data samples that we used for comparison also rely on different methods. For instance, the kinematical data by \citeauthor{Bedregal2006} and \citeauthor{Rawle2013} were derived from stellar absorption lines, whereas \citeauthor{Williams2010} used dynamical models derived from a Navarro-Frenk-White model of the dark matter distribution of their galaxies. Nonetheless, these three studies estimated $\vcirc$ following the approximation described in \citet{Neistein1999}. On the other hand, the data by \citeauthor{Chiu2007}, \citeauthor{Jaffe2014}, and \citeauthor{Davis2016} were based on gas emission lines and they expressed their results in terms of $\vrot$. All these diagnostics are related to the global rotation pattern of the galaxy although they can yield different rotation velocity profiles and, consequently, different locations of the same galaxy onto the TFR \citep{Bertola1995,deBlok2014,Frank2016}. Thus, to make a fairer comparison, we divided our reference samples into two groups, those providing $\vcirc$ measurements and those reporting $\vrot$ values, independently of the tracer used to derive them. We present our results mainly using $\vcirc$ measurements (see Figs.\,\ref{fig:TFVcircBband}, \ref{fig:TFVcircKband}, \ref{fig:vcircmass}, and \ref{fig:sigmavcirc}), but we also include the most representative figures with the $\vrot$ values for comparison (see Figs.\,\ref{fig:TFVrotBband} and \ref{fig:sigmavrot}). Given the goal of our study, we emphasize that this separation does not affect our main results and conclusions.

\section{Results}
\label{secc:results}

In this section, we explore whether the combination of two processes (major mergers and subsequent passive evolution  in the last $\sim 7$\,Gyr) is compatible with the evolution observed in the TFRs of massive S0 galaxies at $z\sim 0.8$ and $z\sim 0$. We built the photometric TF planes in the $B$ and $K$ bands for the S0-like remnants and compared them with the location of real S0 and spiral galaxies at these two redshifts, taken from the literature. Additionally, we analysed the location of our remnants and real galaxies in the stellar TFR, the $\disp$--$M_K$, $\vcirc$--$\disp$, and $\vrot$--$\disp$ planes.

Before proceeding, we caution that our analysis and conclusions are necessarily qualitative. Besides the uncertainties in the corrections mentioned in Sect.~\ref{secc:corrections}, there may be other observational effects that can introduce additional spread in the TF plane, for instance blurring due to the seeing when galaxies at high redshift are observed \citep{Covington2010}. In addition, the kinematical data obtained from different observational methods may differ, and we have included S0s from different sources in the literature to compare with our S0-like remnants (see Sec.~\ref{secc:remarks}).

\subsection{Photometric Tully-Fisher relations}
\label{secc:photoTF}

\subsubsection{$B$-band TFR}
\label{secc:TFRB}

In Fig.~\ref{fig:TFVcircBband} we show the circular velocity as a function of the $B$-band absolute magnitude for all our S0-like remnants (coming from major and minor encounters), compared to real S0 galaxies. In the left panel, we represent the final remnants considering that they are being observed at $z\sim 0.8$. At this redshift we lack observational data for comparison. In the right panel, we show the location of these remnants after $\sim 7$\,Gyr of passive evolution (at $z\sim 0$), compared to local S0 and spiral galaxies by \citet[]{Williams2010a}. In both panels, we represent the TFR fitted to their $z\sim 0$ S0 galaxies by \citeauthor{Williams2010a} as a reference.

Our S0-like remnants resulting from major mergers at $z\sim 0.8$ are offset from the $z \sim 0$ TFR of S0 galaxies (left panel of Fig.~\ref{fig:TFVcircBband}) towards the TFR of the spiral galaxies, but close to the scatter ($\sim0.68-0.88$ mag) reported for the TFR for local S0s \citep[][]{Williams2010a,Bedregal2006}. This is in agreement with previous numerical studies that have shown that major mergers produce remnants that deviate from the local TFR \cite[\eg][]{Covington2010,Tonini2011}. Nonetheless, it is hard to determine whether the $z\sim 0.8$ remnants follow a more `relaxed' TFR towards brighter magnitudes or if they are just outliers of the TFR of local S0s owing to the limited range in magnitude and $\vcirc$ of our sample. The subsequent passive evolution of these S0-like remnants over the last 7\,Gyr moves them onto the local TFR of S0s (right panel of Fig.~\ref{fig:TFVcircBband}). The average decrease in the $B$-band magnitude of $\sim 1.3$\,mag, which corresponds to a decrease of $\sim 70$\% in the $B$-band luminosity, agrees well with the value that would be required to move the TFR of spirals towards the TFR of S0s in the Fornax cluster \citep[offset by $\sim 1.3$--1.7\,mag in the $B$ band; see][]{Bedregal2006}.

Minor mergers have a negligible effect on the TFR (see Fig.~\ref{fig:TFVcircBband}); they just introduce a small dispersion, as already shown by \citet{Qu2010}. The passive evolution of the stellar populations basically moves these remnants to an even better agreement with the local TFR of S0s. Moreover, the  differences in the magnitude evolution are less noticeable because these experiments contained very low gas amounts, so their magnitudes are dominated by the simple fading of their old stellar populations (which are very similar in mass). 

\begin{figure*}[t!h]
\centering
\includegraphics[width=18.3cm]{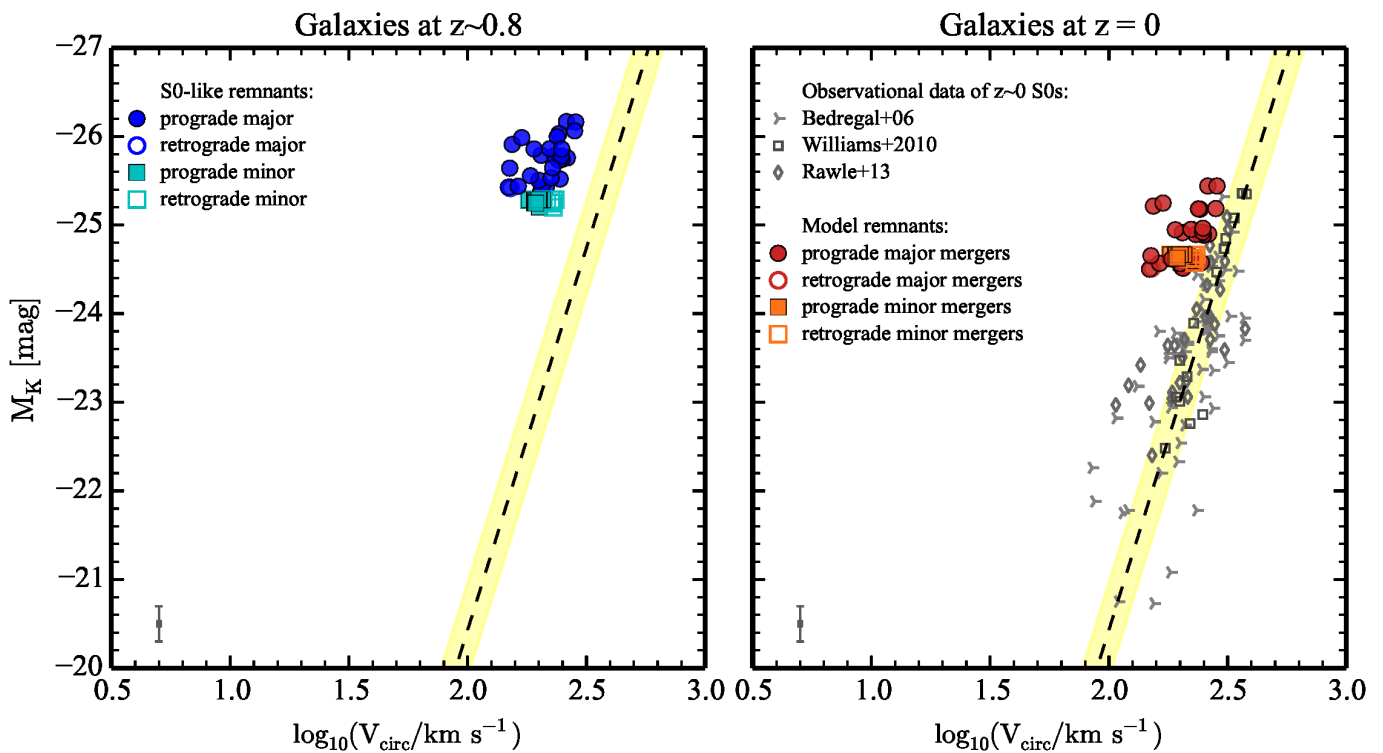}
\caption{\emph{Left panel:} TFR in the $K$ band for the S0-like remnants at $z\sim0.8$, using $\mathrm{\vcirc}$. \emph{Right panel:} local TFR for our remnants at $z\sim 0$ in comparison with local S0 galaxies, again using $\mathrm{\vcirc}$. This figure is analogous to Fig.~\ref{fig:TFVcircBband}, but here we have more samples of real S0 galaxies for the present-day panel. The error bars in the bottom left corner of each panel represents the statistical errors on both axes.}
\label{fig:TFVcircKband}
\end{figure*} 

\subsubsection{$K$-band TFR}
\label{secc:TFRK}

The $K$ band is less sensitive to the effects of recent star formation and dust, so the $K$-band magnitude is a better proxy of the stellar mass of galaxies than the $B$-band magnitude analysed in Sect.~\,\ref{secc:TFRB}. In Fig.\,\ref{fig:TFVcircKband}, we present the $\log(\vcirc)$--$M_K$ plane for our S0-like remnants at $z \sim 0.8$ and at $z\sim 0$ (after $\sim 7$\,Gyr of passive evolution).  We note that we included the effects of the residual star formation in this evolution. We have not found observational data for comparison at $z\sim 0.8$. At $z\sim0$, we have compared our S0-like remnants with the local samples of early-type galaxies by \citet{Bedregal2006}, \citet{Williams2010a}, and \citet{Rawle2013}.

The figure shows that our S0-like remnants at $z\sim 0.8$ are offset from TFR of local S0s (left panel), more clearly than in the $\log(\vcirc)$--$M_B$ plane in Fig.\,\ref{fig:TFVcircBband}, and that they also move towards the local TFR through passive evolution as time goes by. The average change in $M_K$ experienced by the models is such that it moves the S0-like remnants into the $1\sigma$ range of the local TFR of S0s in this band, near the location of bright S0 galaxies. However, the agreement with local data in the $K$ band is not as good as the observed values in the $B$ band. This disagreement is probably related to the hypotheses adopted to transform mass into luminosity in this band, which are model-dependent and have obvious limitations \citep[][]{Bell2001,Tiley2016}. These limitations will be examined in greater detail in a forthcoming paper (Eliche-Moral et al., in prep.).

In contrast with the evolution obtained in the $B$ band, the decrease in luminosity in the $K$ band is much more moderate ($\sim 0.8$\,mag in the last $\sim 7$\,Gyr). It is similar to the offset required in observations to move the TFR of spirals on top of that of S0s in the $K$ band \citep[$\sim 0.5$--0.7\,mag; see][]{denHeijer2015}.
 
In conclusion, the evolutionary scenario tested here for massive S0s (major merger origin at $z\gtrsim 1$ followed by a subsequent passive evolution in isolation) can generate S0s that deviate from the local TFR at $z\sim0.8$, but that later evolve towards the local TRF of S0s at $z\sim 0$.

\subsection{Stellar TFR}\label{secc:stellarTF}

\begin{figure}[t!h]
\centering
   \includegraphics[width=8.8cm]{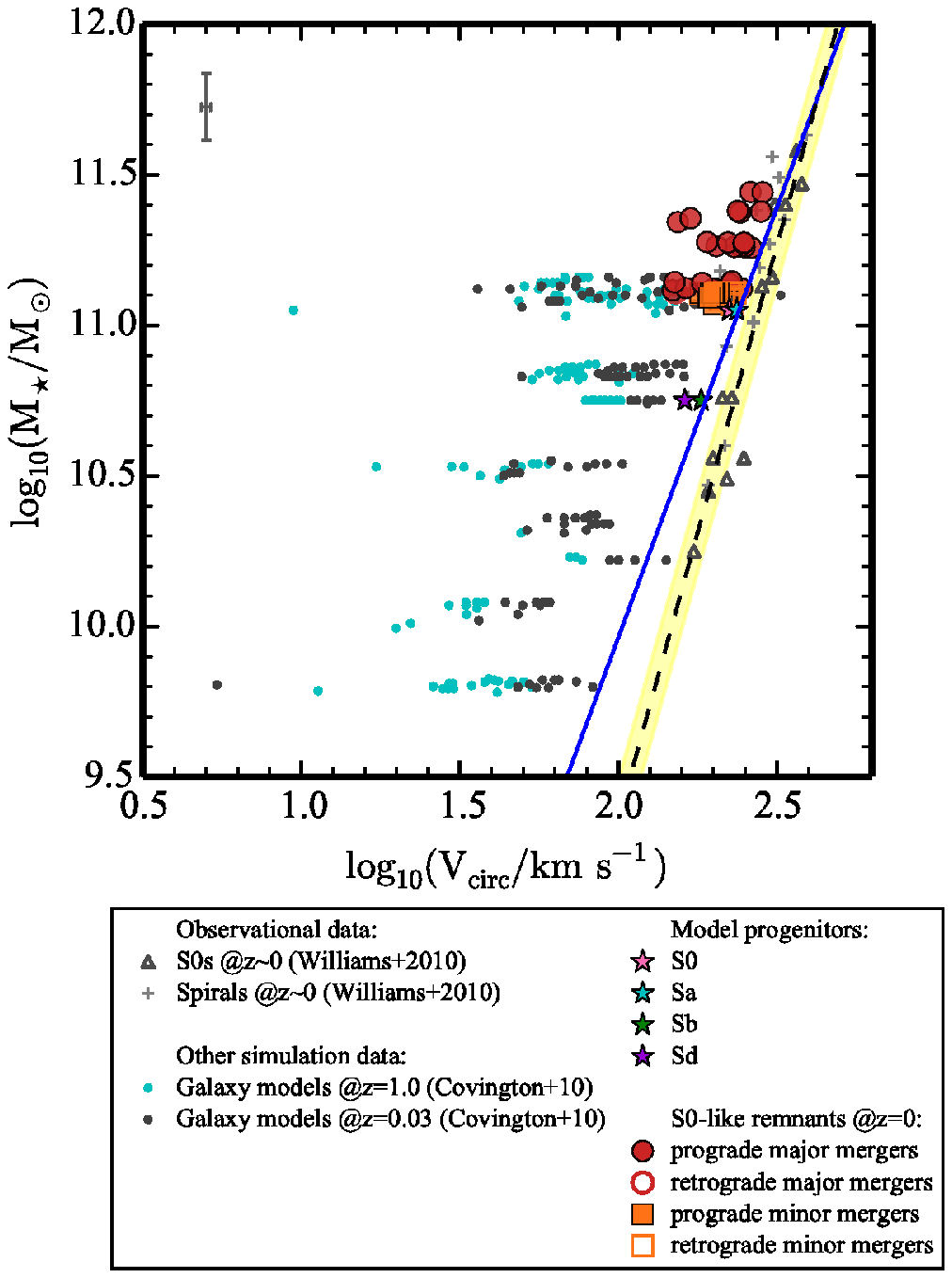}
  \caption{Stellar TFR for the S0-like remnants using $\mathrm{\vcirc}$, compared to the observational data of S0s and spirals by \citet[]{Williams2010a}, and the distribution of mock observations of on-going and remnants of mergers by \citet{Covington2010}. For the latter, the data points correspond to $\vrot$ measurements. We have overplotted the location of the giant progenitors of our merger experiments (gS0, gSa, gSb, and gSd) to show that they satisfy this local relation. We note that the gS0 and gSa models practically overlap in this diagram. The dashed line corresponds to a linear fit to the S0s by \citet[]{Williams2010a} and the yellow band represents its 1$\sigma$ scatter. The blue line denotes the local stellar TFR of spiral galaxies by \citet[]{Williams2010a}. The error bars in the upper left corner correspond to the median of the errors in both axes (see the legend below the figure). }
\label{fig:vcircmass}
\end{figure}

The correlation between $\vcirc$ and the luminosity of a galaxy hides a more `physically-based' correlation between the rotational velocity and the stellar mass of the galaxies \citep[\eg][]{Bell2001,Miller2011}. In Fig.\,\ref{fig:vcircmass} we show the stellar TFR for our S0-like remnants in comparison with the spiral galaxies and local S0s by \citet[]{Williams2010a}. We also represent the mock observations at $z\sim 1$ and $z\sim 0.3$ of merging galaxies simulated by \citet{Covington2010} for illustrative purposes, although they are based on $\vrot$ measurements. We have overplotted the local stellar TFR of spiral galaxies by \citet[]{Williams2010a} and the location in the plane of the progenitor galaxies of our merger experiments to show that they were built to be consistent with this relation.

Our S0-like remnants are slightly displaced from the linear relation drawn by real galaxies in this plane towards slightly lower $\vcirc$ values for a given stellar mass. This behaviour is the same that \citet{Covington2010} found in their merger experiments: mergers always decrease $\vrot$ and increase $\disp$. However, this small disagreement between real and simulated data could be explained by considering the low statistics of the observed sample (which is far from being complete or representative in any sense) as well as the restricted initial conditions of our simulations. In particular, our progenitors reproduce local galaxies in terms of rotation velocities, stellar masses (see their location in Fig.\,\ref{fig:vcircmass}), and gas content \citep{Chilingarian2010}. On the other hand, spirals at $z\gtrsim 1$ rotate slightly faster \citep{Cresci2009} and have higher gas content than the local ones of the same stellar mass \citep{Papovich2005,Genzel2008,Tacconi2008,FoersterSchreiber2009,Law2009}. Both factors could produce remnants with higher levels of rotational support, contributing to a better agreement between S0s resulting from major merger encounters and real S0s in Fig.\,\ref{fig:vcircmass}.

We note that our models cover a narrower stellar mass range compared to observational data because all major merger experiments in GalMer used the same five giant progenitors (all with similar masses). Although the encounters were run with a wide variety of orbits, the final masses of the remnants are  combinations of these five values (the contribution of new stars is always below the initial gas amount available in the progenitors, which is not very high), thus they all are quite similar in the end. Minor mergers consider satellites of mass ratios between 7:1 to 20:1 onto the same gS0 progenitor, hence the masses of their final remnants are very similar as well. 

\subsection{The $\disp$--$M_K$ plane}
\label{secc:sigmaMKplane}

\begin{figure}[t!h]
\centering
   \includegraphics*[width=8.8cm]{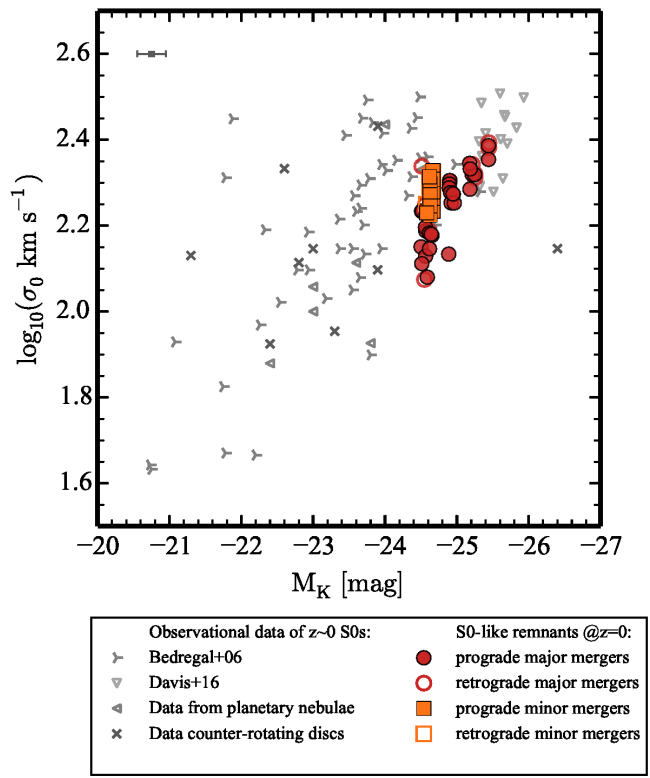} 
  \caption{Comparison between the $\disp$--$M_K$ plane for the S0-like remnants at $z=0$ (after $\sim 7$\,Gyr of passive evolution since $z\sim 0.8$) and observational data of local S0 galaxies. We have distinguished between the remnants resulting from prograde and retrograde encounters. The error bars in the upper left corner represent the statistical errors on both axes (see the legend below the figure).}
\label{fig:sigmaKband}
\end{figure}

In Fig. \ref{fig:sigmaKband} we show the location of our S0-like remnants in the $\disp$--$M_K$ plane in comparison with the distribution of real S0 galaxies at $z\sim0$. Here we include all our S0-like remnants (and not only those with measured circular velocity, see Table\,\ref{tab:kinematicsz0}). We compare with local S0 galaxies by \citet{Bedregal2006}, the gas-rich early-type galaxies by \citet{Davis2016}, and our selection of S0 galaxies hosting counter-rotating discs or with kinematical data derived from the kinematics of their PNe (Table\,\ref{tab:S0scounterrotS0s}). We remark that this is not the Faber-Jackson relation \citep[][]{Faber1976}, which relates $\disp$ with the luminosity of the spheroidal components of galaxies (ellipticals or bulges of disc galaxies). We distinguish between the remnants resulting from direct and retrograde encounters, as in the previous figures. 

Our remnants are in the range $120 < \disp < 220$\, km\,s$^{-1}$ approximately in this plane, values which are very similar to those found in nearby early-type galaxies of similar $K$-band absolute magnitudes. The simulated S0s agree quite well with real S0s, although they tend to be slightly more luminous than present-day S0s of similar $\disp$ values or, conversely, they exhibit slightly lower $\disp$ on average for similar $M_K$ values. In fact, the S0-like remnants are similar to the sample of \citeauthor{Davis2016} in terms of $K$-band luminosity, but their galaxies show higher velocity dispersions. In any case, the difference in magnitudes is small and could be related to the assumptions adopted to convert mass into luminosity in Sect.\,\ref{secc:magnitudes}. We then consider that our S0 remnants formed through major mergers at $z\sim 0.8$ and later passively evolved down to $z\sim 0$ reproduce fairly well the location of present-day S0s with similar $K$-band luminosities in the $\disp$--$M_K$ diagram. 

\subsection{The $\mathrm{\vcirc}$--$\sigma$ plane}
\label{secc:vcircsigmaplane}

\begin{figure}[t!h]
\centering
   \includegraphics[width=9.1cm]{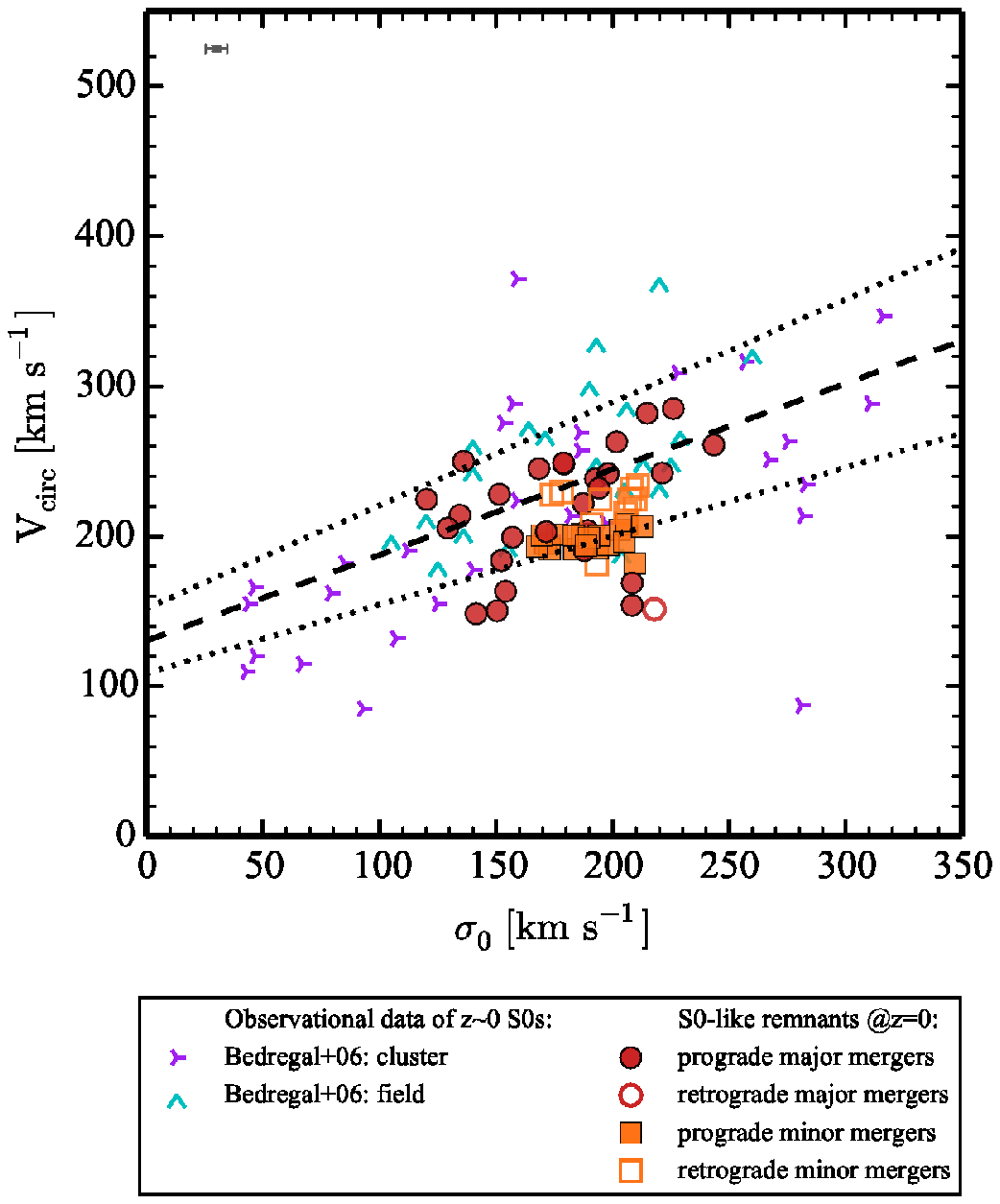}
  \caption{Circular velocity as a function of the central velocity dispersion for our S0-like remnants in comparison with the local S0 galaxies by \protect \citet{Bedregal2006}, distinguishing the environment in which they are located. The dashed line corresponds to a linear fit to the whole sample of local S0s and the dotted lines represent the minimum and maximum slopes according to the errors of the fit. The error bars in the upper left corner correspond to the median of the errors in both axes.}
\label{fig:sigmavcirc}
\end{figure}

In Fig. \ref{fig:sigmavcirc} we show the $\vcirc$--$\disp$ plane for our S0-like remnants and the local S0 galaxies by \citeauthor{Bedregal2006} at $z \sim 0$. For the latter sample, we used different symbols to denote cluster and field galaxies. As observed in previous figures with $\vcirc$, there is not a clear segregation due to the spin-orbit coupling of the remnants (but see Sect.~\ref{secc:TFVrot}). The S0-like remnants reproduce the global kinematics of many real S0 galaxies, especially field galaxies. Thus, our single merger events seem to more consistently represent the typical environmental conditions of field galaxies. The cluster S0s, on the other hand, also populate regions with more extreme velocity dispersions. These differences suggest that the environment certainly affects the kinematics of S0 galaxies and that different mechanisms may have ruled the transformation of spirals into S0s depending on the local environmental density. However, more kinematical data of both real and mock S0 galaxies are needed to establish this notion. In any case, the S0s coming from major mergers reproduce the distribution of many real S0 galaxies in the $\vcirc$--$\disp$ plane as well.

\subsection{Kinematics considering $\vrot$ measurements}
\label{secc:TFVrot}

\begin{figure*}[t!h]
\centering
   \includegraphics[width=18.3cm]{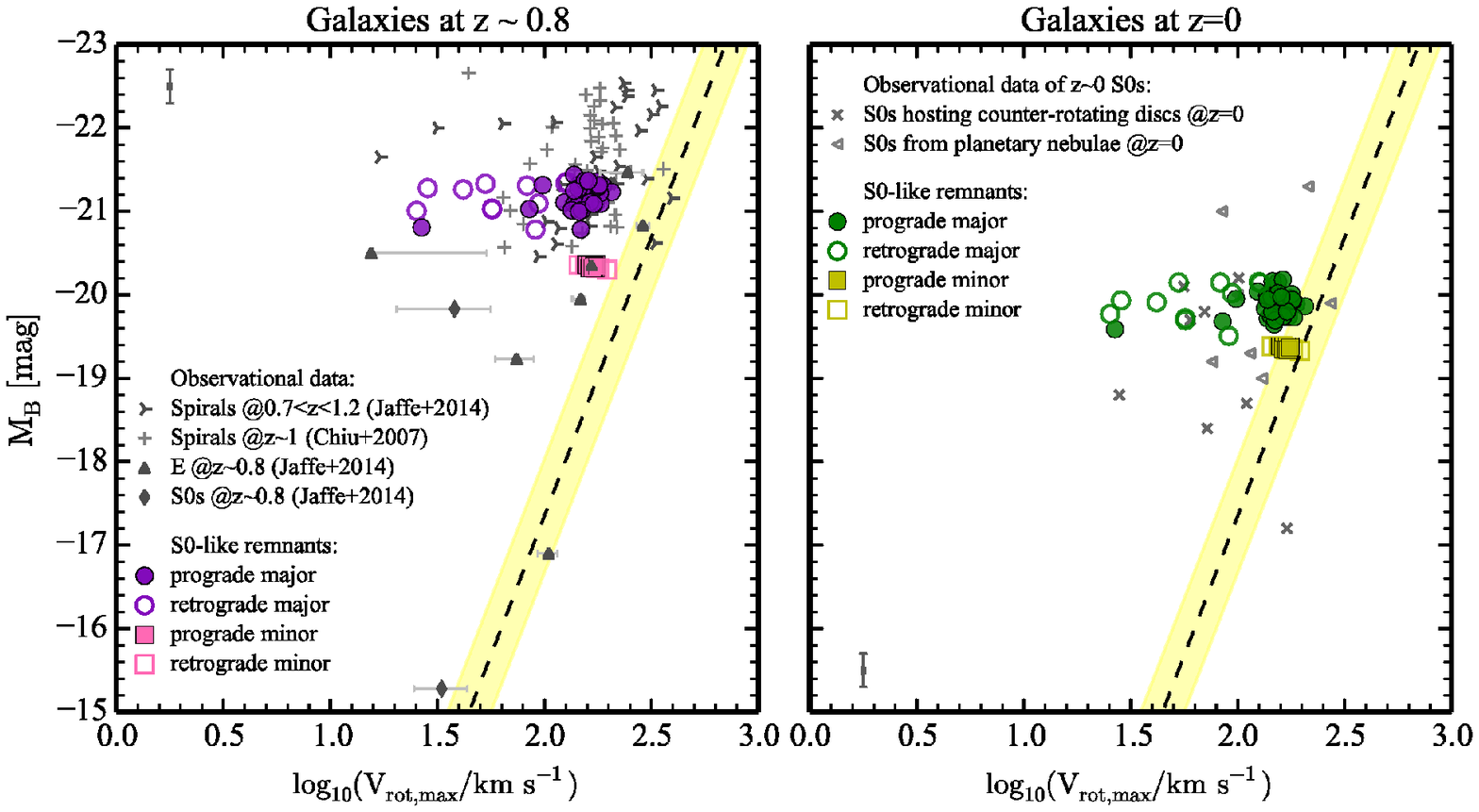}
  \caption{\emph{Left panel:} TFR for our S0-like remnants in the $B$ band at $z\sim 0.8$ using $\vrot$, compared to data of real spirals and S0 galaxies at high redshift. We include the error bars in the kinematics of the early-type galaxies by \citet{Jaffe2014} to highlight the uncertainties of the data at $z\sim 0.8$. \emph{Right panel:} Local TFR for our S0-like remnants at $z \sim 0$ using $\vrot$, in comparison with present-day S0 galaxies (see the legend below the figure). This figure is analogous to Fig.~\ref{fig:TFVcircBband}, but using the maximum rotational velocity measurements instead (see the caption there).}
\label{fig:TFVrotBband}
\end{figure*}

The TFR is also commonly studied using the maximum rotational velocity of galaxies instead of the circular velocity. Therefore, we present the $B$-band TFR for our S0-like remnants in Fig.~\ref{fig:TFVrotBband} (cf. Fig.~\ref{fig:TFVcircBband})  along with real spiral and S0 galaxies. We have kept the fit to the local S0s by \citeauthor{Williams2010a} as a reference, but we caution that the comparison between their fit and our data is not as direct as in Figs.~\ref{fig:TFVcircBband} and \ref{fig:TFVcircKband} because here we use $\vrot$ values instead of $\vcirc$ measurements. Real early-type galaxies at $z\sim 0.8$ seem to be offset from the local TFR of S0s, according to the data by \citet{Jaffe2014}. However, considering their sample size and the fact that it is naturally biased towards galaxies with clearly measurable emission lines, it is hard to tell whether this trend is general or not. More measurements at high-redshift are required to complement these data. A few spirals at $z\sim 1$ by \citeauthor{Chiu2007} and \citeauthor{Jaffe2011} are also scattered towards lower velocity rotation values although they still approximately follow a TFR.

Our S0-like remnants resulting from major mergers at $z\sim 0.8$ significantly deviate from the $z \sim 0$ TFR of S0 galaxies (left panel of Fig.~\ref{fig:TFVrotBband}). The passive evolution of these remnants during 7\,Gyr moves them, especially those resulting from direct encounters, on top of the local TFR of S0s. However, the majority of S0 remnants coming from retrograde major encounters are outliers at $z \sim 0$ (right panel of Fig.~\ref{fig:TFVrotBband}). This segregation is not seen in Fig.~\ref{fig:TFVcircBband} because all major-merger remnants from retrograde orbits (except one) do not have $\vcirc$ measurements available (see Sect.~\ref{secc:calculovcirc}). Additionally, we again corroborate the finding by \citet{Qu2010} that minor mergers have a small effect on the TFR (see the two panels of Fig.~\ref{fig:TFVrotBband}) and only introduce a small scatter in the TFR, as already seen in Fig. \ref{fig:TFVcircBband}.
 
\begin{figure}[t!h]
\centering
   \includegraphics[width=9.1cm]{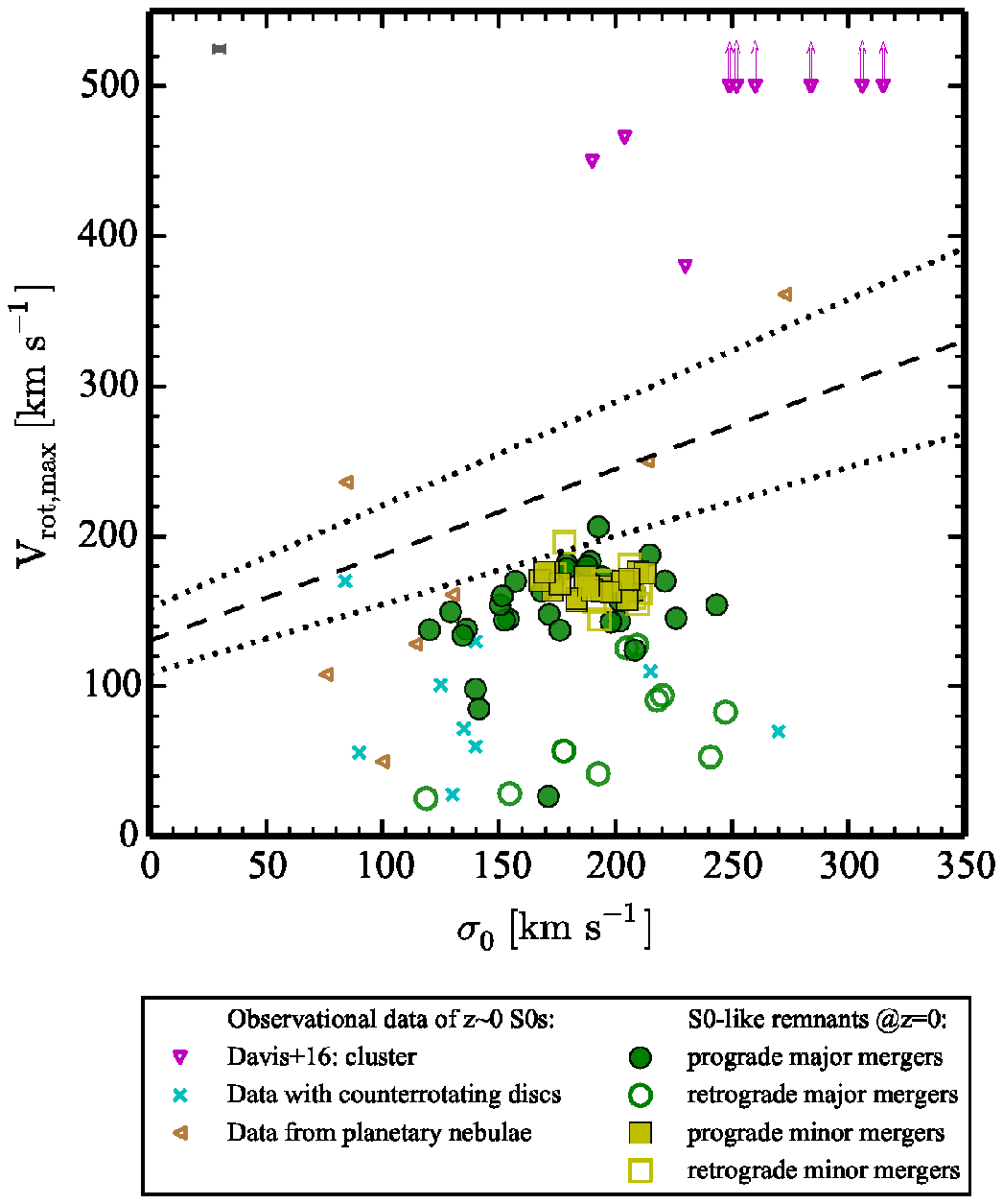}
  \caption{Maximum rotational velocity as a function of the central velocity dispersion for our S0-like remnants in comparison with local early-type galaxies by \protect\citet{Davis2016} and the S0s from Table\,\ref{tab:S0scounterrotS0s}. The magenta triangles with an arrow denote the galaxies by \protect\citeauthor{Davis2016} with rotational velocities higher than the axis limit. The error bars in the upper left corner correspond to the median of the errors in both axes (see the legend below the figure). This figure is analogous to Fig.~\ref{fig:sigmavcirc}, but using the maximum rotational velocity measurements (see the caption there).}
\label{fig:sigmavrot}
\end{figure}

In order to analyse the trend with the spin-orbit coupling in terms of $\vrot$ observed in this figure, we also present the $\vrot$--$\disp$ plane for our S0-like remnants and local S0 galaxies in Fig.~\ref{fig:sigmavrot}. This is analogous to Fig.\,\ref{fig:sigmavcirc}, but using $\vrot$ instead of $\vcirc$. In this case, we compare our models with the gas-rich massive galaxies by \citet{Davis2016}, our selected samples of S0 galaxies hosting counter-rotating discs, and those with kinematical data traced by their PNe (Table\,\ref{tab:S0scounterrotS0s}). We also include the fit indicated in Fig.~\ref{fig:sigmavcirc} as a reference. It is remarkable that our remnants from retrograde major merger encounters, which populate the region with $\mathrm{\vrot < 100\,km\,s^{-1}}$, overlap with many present-day S0s hosting noticeable counter-rotating components, justifying the fact that the galaxies hosting massive counter-rotating components are usually thought to result from major mergers in retrograde orbits.
 
In conclusion, the use of $\vrot$ results in an increase in the scatter of the $B$-band TFR, compared to the use of $\vcirc$ measurements. The S0-like remnants have on average $\sim23\%$ higher $\vcirc$ measurements than their corresponding $\vrot$ values (see Table\,\ref{tab:kinematicsz0}). This may be related to the fact that the calculation of the circular velocity generally relies on data points beyond one disc scalelength, whereas the maximum of the rotational velocity is typically reached at inner radii, thus being affected by the presence of inner discs and counter-rotating components. However, irrespective of which indicator is considered, S0 galaxies at high redshift are clearly displaced from the local TFR of S0s and this offset must be related to their formation process.

\section{Discussion}
\label{secc:discussion}

We have shown that major mergers can produce S0 galaxies at $z\sim 0.8$ that deviate from the local TFR of S0s, and that these galaxies can move towards this relation at $z\sim 0$ by simple passive evolution of their stellar populations. In this section we put our results in the context of other works in the literature. 

Our remnant S0s coming from progenitors with the highest gas content are over-luminous at their early stages as S0 galaxies (at $z\sim 0.8$, as assumed here) for their $\vrot$ value because they host young stellar populations in the centre resulting from merger-induced starbursts that take time to fade enough to reproduce the local TFR (see Figs.\,\ref{fig:TFVcircBband} and \ref{fig:TFVrotBband}). Many authors highlight that a strong starburst at an early phase in the formation of S0 galaxies is required to explain some of their observational characteristics \citep[\eg][]{Christlein2004,Hammer2009}, in particular the detection of young stellar populations in many of them, mostly in their central regions. This supports a merger-origin scenario \citep[][]{Poggianti2001,Yi2005,Kaviraj2007} because cluster-related environmental processes can hardly explain the existence of these starburst episodes without invoking the infall of relevant amounts of gas \citep{Burstein2005,Dekel2006,Cole2014}.

Cluster-related environmental processes can, however, successfully explain the offset of S0s with respect to spirals towards lower luminosities in the TFR at $z\sim 0$ because gas loss is expected to quench the star formation of a spiral (contributing to its fading), but not to alter its kinematics \citep{Bosch2013,Shankar2014}. However, this mechanism cannot explain the S0s outliers at z$\sim$0.8 or the existence of  S0s as bright as spirals. In Figs.\,\ref{fig:TFVcircBband}, \ref{fig:TFVcircKband}, and \,\ref{fig:TFVrotBband}, we have shown that mergers, especially major encounters, are capable of generating S0 galaxies that deviate from the $z=0$ TFR and that can later move towards the local TFR through passive evolution, being as bright as spirals with similar $\vcirc$ or $\vrot$ values.  

There are other mechanisms that can lead to the formation of an S0 galaxy, for instance internal secular processes such as bars or strong gas outflows and inflows induced by torques and nuclear activity \citep{Laurikainen2009,Laurikainen2013,DeRossi2012}. Furthermore, some studies argue that the mechanisms producing S0 galaxies do not mainly depend on the environment, but on the mass of the galaxy \cite[\eg][]{Barway2011,Barway2013,Kraljic2012}. Therefore, mergers and gravitational interactions may be ruling the formation of massive S0s (even of those in clusters), whereas internal secular evolution and environmental processes may be driving this evolution in intermediate-mass to low-mass S0s. 

Although, we find a trend with the environment in the $\vcirc$--$\sigma$ and $\vrot$--$\sigma$ planes, it is not clear whether it induces any significant difference in the TFR. \citet{Torres-Flores2013} studied both the stellar and the baryonic TFRs in the $K$ band for a sample of nearby galaxies in Hickson compact groups. They found that their TFR is very similar to that of field galaxies but with a higher scatter, which they attribute to strong starbursts or active galactic nuclei. On the other hand, \citet{Rawle2013} have found a correlation between the offset in the TFR and the environment for a sample of early-type galaxies in the Coma cluster; the S0s that are located in regions of lower local density are closer to the TFR of spirals and S0s are more abundant towards the centre of the cluster in general. \citet{Puech2010} has directly attributed the scatter in the stellar and baryonic TFRs at $z\sim 0.6$ entirely to mergers. 

All these arguments point to a much more complex formation scenario than just simple fading, bar evolution, or gas stripping for present-day S0 galaxies, favouring a relevant role of mergers in a significant fraction of S0s, at least for the most massive ones. Obviously, all these processes operate at the same time and can enhance or delay the evolution of a spiral into an S0 galaxy. Nevertheless, there are few observational studies that focus on the kinematics of S0s at intermediate to high redshift, due to the difficulties inherent to these sort of studies. More observational effort is necessary in this direction to constrain the formation mechanisms of S0s and the role of environment and mass in them.

\section{Summary and conclusions}
\label{secc:conclusions}

We have explored whether the combination of major-mergers of spiral galaxies and the subsequent passive evolution of the remnant in isolation is a feasible mechanism to explain the location of present-day S0 galaxies in the TFR. To this end, we have studied the kinematics of a set of major and minor merger simulations from the GalMer database that produce S0 remnants with realistic properties. We have measured the maximum rotational velocity and the central velocity dispersion in each remnant mimicking long-slit spectroscopy. We have estimated the effects of passive evolution of their stellar populations since $z\sim 0.8$, accounting for the mixing of the stellar populations in the remnants and the effects of the residual star formation. Then, we have constructed both the photometric and stellar TFRs for our remnants at $z\sim0.8$ and $z\sim0$, considering the two indicators most widely used to characterize the amount of rotation in galaxies, $\vcirc$ and $\vrot$, and qualitatively compared them with observational data.

The main results of our study are the following:

\begin{enumerate}
\item Major mergers of spiral galaxies (initially following the local TFR) give rise to S0-like remnants that are outliers of the photometric and stellar TFRs after $\sim 1$--2\,Gyr of the full merger, in agreement with current data of S0 galaxies at $z\sim 0.8$. On the contrary, minor mergers generate S0-like remnants that just contribute to increasing the dispersion in these relations, in agreement with previous numerical studies.
\item After $\sim 7$\,Gyr of passive evolution in isolation (corresponding to the time period between $z\sim 0.8$ and $z=0$), the S0 remnants resulting from major encounters move towards the local TFR of S0 galaxies, overlapping with it. This passive evolution of the stellar populations in the remnants (including the effects of the residual star formation at $z\sim 0.8$) makes them fade by $\sim 1.3$\,mag and $\sim 0.8$\,mag in the $B$ and $K$ bands during these $\sim 7\,$Gyr, in agreement with the observational offset between the TFR of spirals and S0s in both bands reported by studies on clusters. The passive evolution of the S0 remnants coming from minor mergers for the same time period negligibly changes their initial locations (already verifying the local TFR).
\item The location of the S0-like remnants changes in the TF diagram, depending on whether we use $\vcirc$ or $\vrot$ to study the relation. On average, the $\vcirc$ measurements are $\sim 23\%$ higher than the $\vrot$ values for the same S0 galaxy, which results in lower scatter at high redshift and a much better agreement with the local TFR of S0 galaxies when $\vcirc$ measurements are used. However, the behaviour over time found in both cases is practically the same.
\item The S0 remnants at $z\sim 0$ are also consistent with the distribution of local S0 galaxies in the $\disp$--$M_K$, $\vcirc$--$\sigma$, and $\vrot$--$\sigma$ planes. There is a segregation due to the spin-orbit coupling of the S0-like remnants in the $\vrot$--$\sigma$ plane and in the TFR when using $\vrot$. The remnants of retrograde major mergers overlap with some of the local S0s that host counter-rotating discs, supporting the merger origin traditionally assumed for these systems. 
\end{enumerate}

Therefore, we conclude that the combination of the two mechanisms proposed here (major merger followed by passive evolution in relative isolation) is a plausible scenario for the formation and evolution of massive S0s because it can explain why these galaxies deviate from the local TFR of S0 galaxies at $z\sim 0.8$ and have later evolved towards it at $z=0$. Moreover, these mechanisms can also explain the offset or high level of scatter observed in this relation for S0s. If the evolution of many massive S0s follows this scenario, future studies should find a high number of massive S0 galaxies outlying the local TFR at high redshifts.

\small  
%
\begin{acknowledgements}    We thank the anonymous referee for the constructive comments and suggestions that significantly improved the manuscript. The authors would like to acknowledge I. Chilingarian, P. Di Matteo, F. Combes, A.~L. Melchior, and B. Semelin for creating the GalMer database. TT acknowledges the financial support of the Consejo Nacional de Ciencia y Tecnolog\'ia (CONACyT) under the program ``Estancias Posdoctorales Vinculadas al Fortalecimiento de la Calidad del Posgrado Nacional'' and the hospitality of Thomas M\"{u}ller and V\'{i}ctor Al\'{i} Lagoa. Supported by the Spanish Ministry of Economy and Competitiveness under project AYA2012-31277. HA is grateful for the financial support of CONACyT Project No. 179662 and UNAM-PAPIIT IN108914. We acknowledge the use of the HyperLeda database (http://leda.univ-lyon1.fr).  This research has made use of the NASA/IPAC Extragalactic Database (NED) which is operated by the Jet Propulsion Laboratory, California Institute of Technology, under contract with the National Aeronautics and Space Administration. This publication makes use of data products from the Two Micron All Sky Survey, which is a joint project of the University of Massachusetts and the Infrared Processing and Analysis Center/California Institute of Technology, funded by the National Aeronautics and Space Administration and the National Science Foundation. 
\end{acknowledgements}

\bibliographystyle{aa}
\bibliography{massiveellipticals}

\begin{appendix}
\section{Dust extinction}
\label{appendix:dust}

Given that S0s have traditionally been considered dust-free systems, most studies analysing the TFR of lenticulars do not apply any dust extinction corrections to their magnitudes, even in optical bands \citep[see \eg][]{Bedregal2006,Davis2011}. Nevertheless, many recent works claim that the dust content of S0s may have been significantly underestimated \citep{Patil2007,Riad2010,Rowlands2012,diSerego2013} and that an improper dust correction could lead to underestimation of the luminosity of a galaxy (or, equivalently, its stellar mass) to $\sim 40$\% of its real value \citep{Zibetti2009}. 

Combining the extinction measurements made by \citet{Finkelman2010} and  \citet{Annibali2010} for nearby S0s, the average extinction ($A$) of an E-S0 galaxy is $A_B\sim 1.2$\,mag and $A_K\sim 0.15$\,mag in the $B$ and $K$ bands, respectively. However, this extinction can be as high as $A_B\sim 7$\,mag for some of their local objects, which is much higher than the $A_B\lesssim 0.3$\,mag typically predicted for local galaxies \citep[see \eg][]{Falcon-Barroso2006,Williams2010a}. This effect might be more dramatic as we move towards higher redshifts, because S0s tend to be bluer \citep{Mei2006,Huertas-Company2010}. Moreover, real major-merger remnants are strongly extincted by dust \citep{Bekki2000,Bekki2001b,Rothberg2004,Yuan2012}. Therefore, the magnitudes of S0s in observational TFRs probably include a significant dust extinction term which has not been properly corrected in most studies. 

Unfortunately, a proper implementation of dust effects in galaxy simulations is difficult because it depends on poorly known properties of galaxies (dust/stars geometry, extinction curve, a full radiative transfer model for realistic predictions, etc.), requires hypotheses about the stellar population synthesis models that are equally uncertain, and uses recipes that frequently entail serious numerical problems \citep[see \eg][]{Jonsson2010,Bekki2013a,Bekki2013b}. To avoid all these uncertainties, we have opted for a simplified approach and accounted for this effect in a global manner by applying the average dust extinctions of local S0s in the $B$ and $K$ bands to all our S0-like remnants at all redshifts, \ie\, $A_B\sim 1.2$\,mag and $A_K\sim 0.15$. We note that these dust extinction values are probably underestimating the real effects of dust in major mergers, especially at the early post-merger stages, because they are based on studies of low-redshift S0s, which are gas-poor and already passive. Thus, our dust correction can be considered as a lower limit, especially at high redshift. 

\section{Weighting of the kinematical data}
\label{appendix:weight}

Traditionally, N-body studies derive mass-weighted kinematical diagnostics \citep[see][among others]{Eliche-Moral2006,Gonzalez-Garcia2006a,Jesseit2009,Bois2010,Bois2011}, but the kinematical data in real galaxies are generally obtained from stellar absorption lines or gas emission lines \citep[\eg][]{Pizzella2004,Silchenko2011,Fabricius2012}. Thus, to ensure a fair comparison with real S0 galaxies, we compared the values of $\vrot$ and $\disp$ obtained with the traditional mass-weighted kinematical diagnostic to those based on $V$-band luminosity-weighted kinematics of the particles.%

\begin{figure}[t!h]
\centering
   \includegraphics[width=8.0cm]{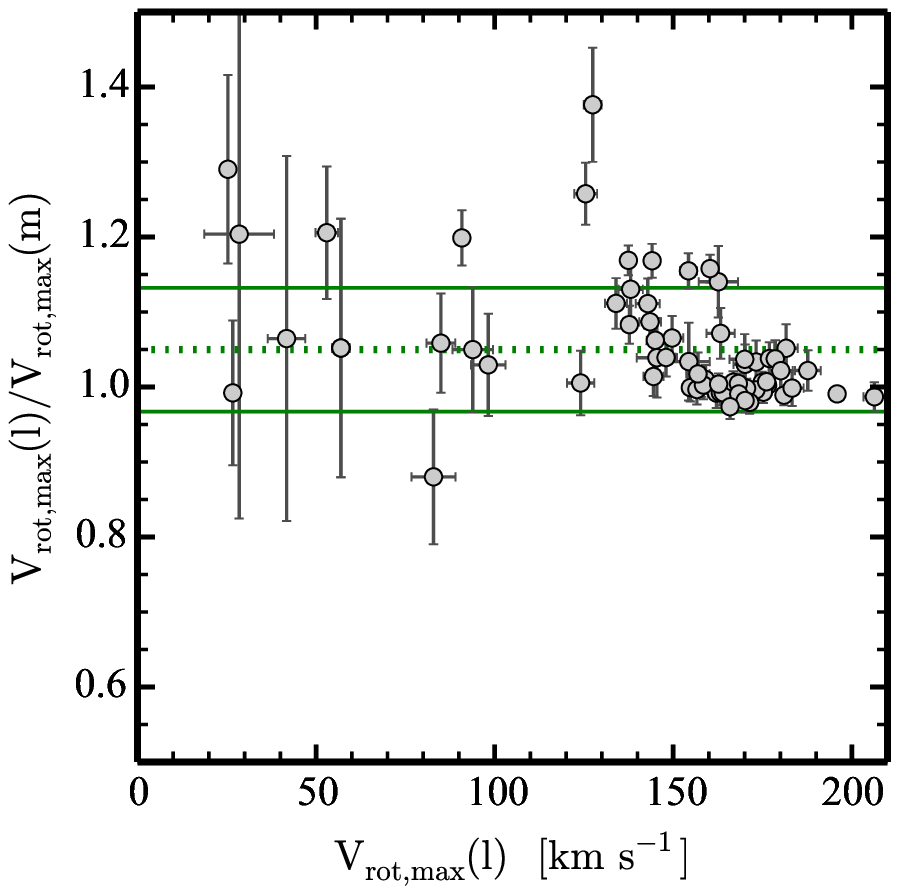}
  \caption{Ratio between the values of the maximum rotation velocity obtained when weighting by the luminosity ($\vrot(\mathrm{l})$) and by the mass ($\vrot(\mathrm{m})$) as a function of $\vrot(\mathrm{l})$ for our S0-like remnants (grey circles). The green dotted line corresponds to the sample mean, and the green solid lines represent the  $1\sigma$ scatter. The error bars correspond to the individual errors in each axis.}
\label{fig:ratiovrot}
\end{figure} 

In Fig.~\ref{fig:ratiovrot} we show the ratio between the maximum rotation velocities of the data weighted by the luminosity ($\vrot(\mathrm{l})$) and by the mass ($\vrot(\mathrm{m})$) of the particles as a function of $\vrot(\mathrm{l})$. In general, the results are within $\sim 20\%$, but there is a higher scatter in the remnant models with the lowest rotational velocities for which the difference can be as high as $\sim 40\%$. Some of these cases correspond to remnants of retrograde orbits, which we discuss in the main text. At any rate, the average maximum rotational velocities found from both methods are similar, namely $145 \pm 42$\,km\,s$^{-1}$ for the luminosity-weighted data and $140 \pm 43$\,km\,s$^{-1}$ for the mass-weighted data.

\begin{figure}[t!h]
\centering
   \includegraphics[width=8.0cm]{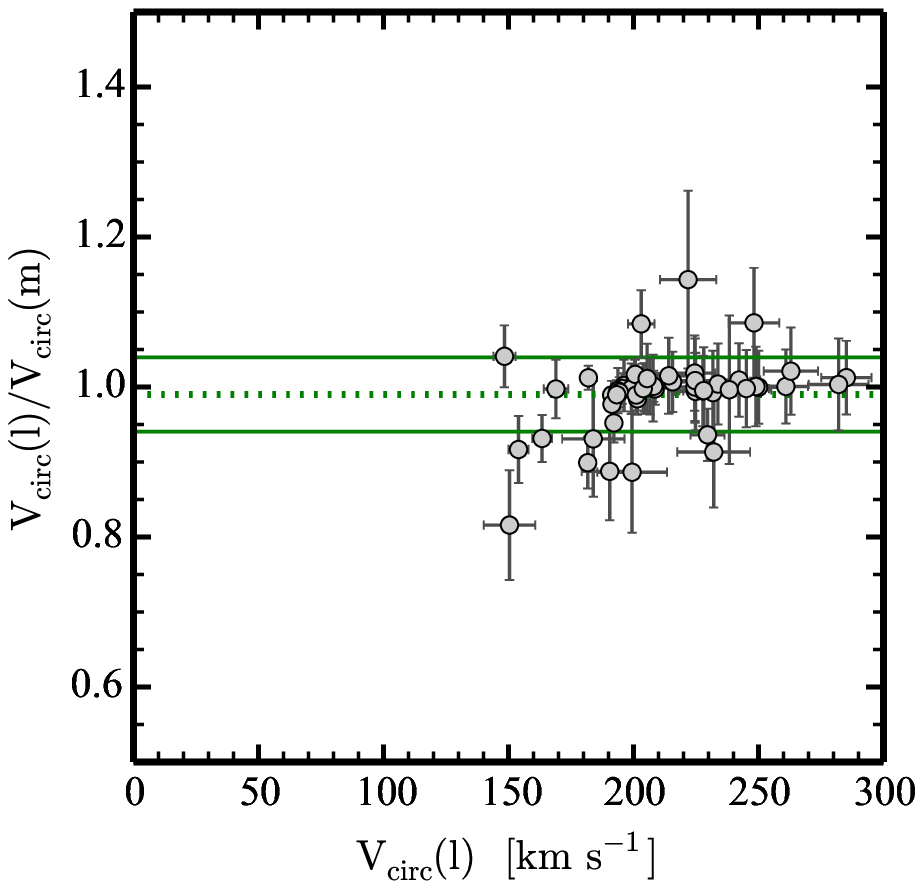}
  \caption{Ratio between the circular velocity measurements obtained when weighting by the luminosity ($\vcirc(\mathrm{l})$) and by the mass ($\vcirc(\mathrm{m})$) as a function of $\vcirc(\mathrm{l})$ for our S0-like remnants (grey circles). The green dotted line corresponds to the sample mean, and the green solid lines represent the $1\sigma$ scatter. The error bars correspond to the individual errors in each axis.}
\label{fig:ratiovcirc}
\end{figure} 

Likewise, in Fig.\,~\ref{fig:ratiovcirc} we present the ratio between the circular velocities of both weighting methods ($\vcirc(\mathrm{l})/\vcirc(\mathrm{m})$) as a function of the luminosity-weighted circular velocity ($\vcirc(\mathrm{l})$). The models show a higher rotation when the circular velocity is considered compared to the rotation velocity measurements in  Fig.~\ref{fig:ratiovrot}. The average circular velocities of both weighting methods are very similar, $169 \pm 88$\,km\,s$^{-1}$ for the luminosity weighted and $165 \pm  92$\,km\,s$^{-1}$  for the mass weighted. In this case, the mean ratio is $\sim 1$. The weighting method thus barely changes the circular velocity measurements.

\begin{figure}[t!h]
\centering
 \includegraphics[width=8.0cm]{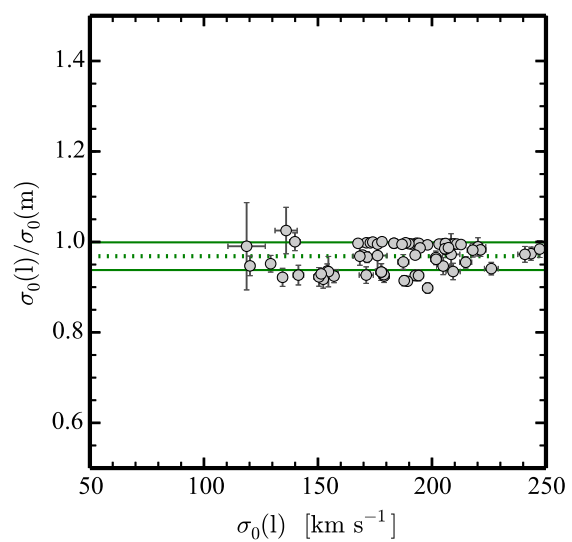}
  \caption{Ratio of the luminosity- and mass-weighted values of the central velocity dispersion $(\disp(\mathrm{l})/\disp(\mathrm{m}))$ of our S0-like remnants as a function of the luminosity-weighted central velocity dispersion $(\disp(\mathrm{l}))$. See the caption of Fig. \ref{fig:ratiovrot}.}
\label{fig:ratiosigma}
\end{figure}

In Fig. \ref{fig:ratiosigma}, we have compared the luminosity- and mass-weighted central velocity dispersions, $\disp(\mathrm{l})$ and $\disp(\mathrm{m})$. These values agree more closely, but the $\disp(\mathrm{l})$ are in general lower than the corresponding $\disp(\mathrm{m})$ estimates (mean of $186.5 \pm 27.7$\,km\,s$^{-1}$ and $192.6 \pm 27.7$\,km\,s$^{-1}$, respectively). This occurs because hybrid particles tend to end up in the central regions of the remnants following circular orbits and hence have a lower dispersion than the mean rotation pattern in the centre. Given that they trace the young stellar populations that form during the merger, they are usually brighter and bluer than the old stellar particles in that region. Consequently, in spite of their mass, they contribute too much to the moments of the LOSVD in that region when the luminosity weight is used to estimate $\disp$.

\begin{figure}[t!h]
\centering
   \includegraphics[width=8.0cm]{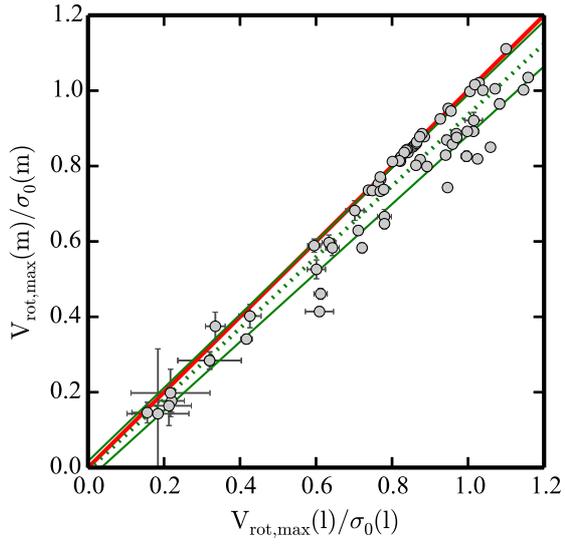}
  \caption{Mass-weighted $\vrot/\disp$ ratio of the S0-like remnants compared with the luminosity-weighted $\vrot/\disp$ ratio. The green dotted line corresponds to the linear fit performed to the data and the green solid lines represent the minimum and maximum slope according to the errors of the fit. The red line denotes the 1:1 relation.}
\label{fig:ratiosvrotsigma}
\end{figure}

Finally, in Fig.\,~\ref{fig:ratiosvrotsigma} we have plotted the $\mathbf{\vrot/\disp}$ values derived from both methods to show that they are consistent, although the luminosity-weighted values are systematically slightly higher.

We carried out our analysis with both methods and, for our purposes of establishing a qualitative comparison in the TF plane, the particular choice of weighting quantity does not have an impact on our conclusions. However, given that the kinematical diagnostics in real galaxies are mostly obtained from luminosity-dependent techniques, a luminosity-weighted kinematics for the simulated data is more likely to capture this dependency and hence we decided to present only this case in Sect.~\ref{secc:results}.

\section{Table with the kinematical data of the S0-like remnants}
\label{appendix:finaltable}
\onecolumn
\begin{longtab}
\begin{landscape}
{\footnotesize
\begin{center}
\begin{longtable}{llccccccccccccc}
\caption{Mass, kinematical parameters (weighted by the luminosity in the V band and by the stellar content of the particles), and magnitudes (at high redshift and at present day) for the S0 remnants resulting from GalMer major and minor merger simulations.}\label{tab:kinematicsz0}\\\\\vspace{-0.5cm}\\
\hline
\hline
&\multicolumn{1}{c}{} & & \multicolumn{3}{c}{luminosity weighted} &&\multicolumn{3}{c}{mass weighted}&\multicolumn{2}{c}{}& \multicolumn{2}{c}{}\\\cline{4-6}\cline{8-10}
&\multicolumn{1}{c}{} & &\multicolumn{3}{c}{}& &\multicolumn{3}{c}{}&\multicolumn{2}{c}{}& \multicolumn{2}{c}{}\\
&\multicolumn{1}{c}{model} & \multicolumn{1}{c}{$\log(\Ms)$} & \multicolumn{1}{c}{$\vcirc$} & \multicolumn{1}{c}{$\disp$} & \multicolumn{1}{c}{$\vrot$} & & \multicolumn{1}{c}{$\vcirc$} & \multicolumn{1}{c}{$\disp$} & \multicolumn{1}{c}{$\vrot$}&\multicolumn{1}{c}{$M_{B}(z=0.8)$}&\multicolumn{1}{c}{$M_{K}(z=0.8)$}&\multicolumn{1}{c}{$M_{B}(z=0)$}&\multicolumn{1}{c}{$M_{K}(z=0)$} \\
&   &  [$\Msun$] & [km\,s$^{-1}$] & [km\,s$^{-1}$]& [km\,s$^{-1}$] && [km\,s$^{-1}$] & [km\,s$^{-1}$]& [km\,s$^{-1}$] & [mag] & [mag] & [mag] & [mag]\\
\multicolumn{1}{c}{(1)} & \multicolumn{1}{c}{(2)} & \multicolumn{1}{c}{(3)} & \multicolumn{1}{c}{(4)} & \multicolumn{1}{c}{(5)} & \multicolumn{1}{c}{(6)} && \multicolumn{1}{c}{(7)}&\multicolumn{1}{c}{(8)} &\multicolumn{1}{c}{(9)}&\multicolumn{1}{c}{(10)}&\multicolumn{1}{c}{(11)}&\multicolumn{1}{c}{(12)}&\multicolumn{1}{c}{(13)}\\\hline
\endfirsthead
\\
\\
\\
\caption*{\\\emph{(Continued)}}
\\ 
\hline \hline
&\multicolumn{1}{c}{} & &\multicolumn{3}{c}{luminosity weighted} &&\multicolumn{3}{c}{mass weighted}&\multicolumn{2}{c}{}& \multicolumn{2}{c}{}\\\cline{4-6}\cline{8-10}
&\multicolumn{1}{c}{} & &\multicolumn{3}{c}{}&&\multicolumn{3}{c}{}&\multicolumn{2}{c}{}& \multicolumn{2}{c}{}\\
&\multicolumn{1}{c}{model} & \multicolumn{1}{c}{$\Ms$} & \multicolumn{1}{c}{$\vcirc$} & \multicolumn{1}{c}{$\disp$} & \multicolumn{1}{c}{$\vrot$} && \multicolumn{1}{c}{$\vcirc$} & \multicolumn{1}{c}{$\disp$} & \multicolumn{1}{c}{$\vrot$}&\multicolumn{1}{c}{$M_{B}(z=0.8)$}&\multicolumn{1}{c}{$M_{K}(z=0.8)$}&\multicolumn{1}{c}{$M_{B}(z=0)$}&\multicolumn{1}{c}{$M_{K}(z=0)$}  \\
&   &  [$\Msun$] & [km\,s$^{-1}$] & [km\,s$^{-1}$]& [km\,s$^{-1}$] && [km\,s$^{-1}$] & [km\,s$^{-1}$]& [km\,s$^{-1}$] & [mag] & [mag] & [mag] & [mag]\\
\multicolumn{1}{c}{(1)} & \multicolumn{1}{c}{(2)} & \multicolumn{1}{c}{(3)} & \multicolumn{1}{c}{(4)} & \multicolumn{1}{c}{(5)} & \multicolumn{1}{c}{(6)} && \multicolumn{1}{c}{(7)}&\multicolumn{1}{c}{(8)} &\multicolumn{1}{c}{(9)}&\multicolumn{1}{c}{(10)}&\multicolumn{1}{c}{(11)}&\multicolumn{1}{c}{(12)}&\multicolumn{1}{c}{(13)}\\\hline
\endhead
& gE0 & 11.20 & $\ldots$ & 205.2 $\pm$ 4.2   & 23  $\pm$ 9  1  & &  $\ldots$ & 206.0 $\pm$ 4.9 & 12  $\pm$ 4   & -20.9 & -25.1& $\ldots$ &$\ldots$\\
& gS0 & 11.05 & 240 $\pm$ 2  & 103.2 $\pm$ 0.6  & 250 $\pm$ 0.3  && 244 $\pm$ 3  & 103.8 $\pm$ 0.5 & 251 $\pm$ 0.4 & -20.5 & -24.7&$\ldots$&$\ldots$\\
& gSa & 11.05 & 252 $\pm$ 3 & 100.0 $\pm$ 0.1  & 259 $\pm$ 0.5  &&  247 $\pm$ 2    & 97.9 $\pm$ 0.1 & 257 $\pm$ 1 & -21.7 & -25.2&$\ldots$&$\ldots$\\
& gSb & 10.75 & 194 $\pm$2 & 77.7  $\pm$ 0.8  & 197 $\pm$ 0.4  &&  192 $\pm$ 2 & 78.8  $\pm$ 0.8 & 194 $\pm$ 0.4 & -21.6 & -24.6&$\ldots$&$\ldots$\\
& gSd & 10.75 & 174 $\pm$ 2 & 50.5  $\pm$ 0.7  & 171 $\pm$ 0.4  &&  174 $\pm$ 2 & 51.2  $\pm$ 0.6 & 169 $\pm$ 0.4 & -22.3 & -24.8&$\ldots$&$\ldots$ \\\hline
1 & gE0gSao1$^{\dagger}$     &  11.44 &    261 $\pm$      9 &  243.4 $\pm$    3.4 &    154 $\pm$      6 &&    261 $\pm$      9 &  249.9 $\pm$    0.9 &    149 $\pm$      5 &  -21.3 &  -26.2 &  -20.2 &   -25.4\\ 
2 & gE0gSao16    &  11.44 &      $\ldots$ &  240.8 $\pm$    3.1 &     53 $\pm$      3 &&      $\ldots$ &  247.6 $\pm$    2.9 &     44 $\pm$      3 &  -21.3 &  -26.2 &  -20.1 &   -25.4\\ 
3 & gE0gSao44    &  11.44 &      $\ldots$ &  247.2 $\pm$    3.4 &     83 $\pm$      6 &&      $\ldots$ &  251.2 $\pm$    3.1 &     94 $\pm$      5 &  -21.3 &  -26.2 &  -20.1 &   -25.4\\ 
4 & gE0gSao5$^{\dagger}$     &  11.44 &    285 $\pm$     10 &  226.0 $\pm$    3.0 &    145 $\pm$      6 &&    282 $\pm$     10 &  240.2 $\pm$    0.9 &    140 $\pm$      5 &  -21.3 &  -26.2 &  -20.2 &   -25.4\\ 
5 & gE0gSbo44    &  11.34 &      $\ldots$ &  220.2 $\pm$    3.4 &     94 $\pm$      6 &&      $\ldots$ &  222.6 $\pm$    2.8 &     89 $\pm$      5 &  -21.1 &  -25.9 &  -20.0 &   -25.2\\ 
6 & gE0gSbo5     &  11.34 &    154 $\pm$      4 &  208.4 $\pm$    3.6 &    124 $\pm$      4 &&    168 $\pm$      6 &  209.5 $\pm$    3.3 &    123 $\pm$      4 &  -21.1 &  -25.9 &  -20.0 &   -25.2\\ 
7 & gE0gSdo16    &  11.36 &      $\ldots$ &  209.3 $\pm$    3.1 &    127 $\pm$      2 &&      $\ldots$ &  223.9 $\pm$    2.3 &     93 $\pm$      7 &  -21.4 &  -26.0 &  -20.1 &   -25.3\\ 
8 & gE0gSdo44    &  11.36 &      $\ldots$ &  204.9 $\pm$    2.9 &    125 $\pm$      3 &&      $\ldots$ &  216.5 $\pm$    2.7 &    100 $\pm$      3 &  -21.3 &  -26.0 &  -20.2 &   -25.3\\ 
9 & gE0gSdo5$^{\dagger}$     &  11.36 &    169 $\pm$      5 &  208.4 $\pm$    3.8 &    163 $\pm$      5 &&    169 $\pm$      4 &  214.3 $\pm$    3.6 &    143 $\pm$      5 &  -21.3 &  -26.0 &  -20.2 &   -25.2\\ 
10& gS0dE0o100$^{\dagger}$   &  11.11 &    196 $\pm$      5 &  205.0 $\pm$    0.7 &    158 $\pm$      2 &&    196 $\pm$      5 &  205.8 $\pm$    0.7 &    158 $\pm$      2 &  -20.4 &  -25.3 &  -19.4 &   -24.7\\ 
11& gS0dE0o101   &  11.11 &    182 $\pm$      2 &  209.6 $\pm$    0.7 &    176 $\pm$      2 &&    180 $\pm$      2 &  210.7 $\pm$    0.7 &    176 $\pm$      2 &  -20.4 &  -25.3 &  -19.4 &   -24.7\\ 
12& gS0dE0o102   &  11.11 &    191 $\pm$      3 &  183.4 $\pm$    0.8 &    157 $\pm$      2 &&    193 $\pm$      3 &  184.0 $\pm$    0.7 &    156 $\pm$      2 &  -20.4 &  -25.3 &  -19.4 &   -24.7\\ 
13& gS0dE0o103$^{\dagger}$   &  11.11 &    232 $\pm$      9 &  208.1 $\pm$    0.7 &    159 $\pm$      2 &&    234 $\pm$      9 &  209.2 $\pm$    0.7 &    157 $\pm$      2 &  -20.4 &  -25.3 &  -19.4 &   -24.7\\ 
14& gS0dE0o104$^{\dagger}$   &  11.11 &    225 $\pm$      8 &  209.6 $\pm$    0.7 &    155 $\pm$      2 &&    226 $\pm$      8 &  210.6 $\pm$    0.7 &    155 $\pm$      2 &  -20.4 &  -25.3 &  -19.4 &   -24.7\\ 
15& gS0dE0o105$^{\dagger}$   &  11.11 &    234 $\pm$      9 &  210.9 $\pm$    0.7 &    162 $\pm$      2 &&    233 $\pm$      9 &  211.9 $\pm$    0.7 &    163 $\pm$      2 &  -20.4 &  -25.3 &  -19.4 &   -24.7\\ 
16& gS0dE0o106   &  11.11 &    182 $\pm$      2 &  193.2 $\pm$    0.8 &    144 $\pm$      3 &&    202 $\pm$      6 &  193.7 $\pm$    0.7 &    142 $\pm$      2 &  -20.4 &  -25.3 &  -19.4 &   -24.7\\ 
17& gS0dE0o109   &  11.11 &    195 $\pm$      3 &  171.2 $\pm$    0.8 &    176 $\pm$      2 &&    196 $\pm$      3 &  171.6 $\pm$    0.7 &    175 $\pm$      2 &  -20.4 &  -25.3 &  -19.4 &   -24.7\\ 
18& gS0dE0o110$^{\dagger}$   &  11.11 &    208 $\pm$      7 &  191.0 $\pm$    0.8 &    157 $\pm$      2 &&    208 $\pm$      7 &  191.9 $\pm$    0.7 &    157 $\pm$      2 &  -20.4 &  -25.3 &  -19.4 &   -24.7\\ 
19& gS0dE0o111   &  11.11 &    192 $\pm$      3 &  193.7 $\pm$    0.8 &    167 $\pm$      2 &&    202 $\pm$      4 &  195.2 $\pm$    0.7 &    167 $\pm$      2 &  -20.4 &  -25.3 &  -19.4 &   -24.7\\ 
20& gS0dE0o113$^{\dagger}$   &  11.11 &    207 $\pm$      5 &  212.7 $\pm$    0.7 &    175 $\pm$      2 &&    207 $\pm$      5 &  214.0 $\pm$    0.7 &    176 $\pm$      2 &  -20.4 &  -25.3 &  -19.4 &   -24.7\\ 
21& gS0dE0o115$^{\dagger}$   &  11.11 &    200 $\pm$      4 &  183.3 $\pm$    0.8 &    158 $\pm$      2 &&    199 $\pm$      4 &  183.8 $\pm$    0.7 &    158 $\pm$      2 &  -20.4 &  -25.3 &  -19.4 &   -24.7\\ 
22& gS0dE0o117$^{\dagger}$   &  11.11 &    200 $\pm$      4 &  194.2 $\pm$    0.7 &    163 $\pm$      2 &&    200 $\pm$      4 &  194.8 $\pm$    0.7 &    164 $\pm$      2 &  -20.4 &  -25.3 &  -19.4 &   -24.7\\ 
23& gS0dE0o98$^{\dagger}$    &  11.11 &    199 $\pm$      5 &  189.9 $\pm$    0.8 &    164 $\pm$      2 &&    200 $\pm$      5 &  190.6 $\pm$    0.7 &    165 $\pm$      2 &  -20.4 &  -25.3 &  -19.4 &   -24.7\\ 
24& gS0dE0o99    &  11.11 &    204 $\pm$      3 &  203.1 $\pm$    0.7 &    170 $\pm$      2 &&    204 $\pm$      6 &  204.1 $\pm$    0.7 &    172 $\pm$      2 &  -20.4 &  -25.3 &  -19.4 &   -24.7\\ 
25& gS0dS0o100   &  11.09 &    191 $\pm$      2 &  173.0 $\pm$    0.8 &    164 $\pm$      2 &&    196 $\pm$      3 &  173.3 $\pm$    0.7 &    165 $\pm$      2 &  -20.3 &  -25.2 &  -19.3 &   -24.6\\ 
26& gS0dS0o101$^{\dagger}$   &  11.09 &    194 $\pm$      4 &  188.4 $\pm$    0.8 &    167 $\pm$      2 &&    195 $\pm$      4 &  188.8 $\pm$    0.7 &    166 $\pm$      2 &  -20.3 &  -25.2 &  -19.3 &   -24.6\\ 
27& gS0dS0o102   &  11.09 &    194 $\pm$      3 &  198.1 $\pm$    0.7 &    163 $\pm$      2 &&    196 $\pm$      3 &  199.3 $\pm$    0.7 &    162 $\pm$      2 &  -20.3 &  -25.2 &  -19.3 &   -24.6\\ 
28& gS0dS0o103$^{\dagger}$   &  11.09 &    228 $\pm$      8 &  174.1 $\pm$    0.8 &    175 $\pm$      2 &&    229 $\pm$      7 &  174.1 $\pm$    0.7 &    174 $\pm$      2 &  -20.3 &  -25.2 &  -19.3 &   -24.6\\ 
29& gS0dS0o105   &  11.09 &    209 $\pm$      3 &  205.9 $\pm$    0.7 &    181 $\pm$      2 &&    209 $\pm$      3 &  206.7 $\pm$    0.7 &    183 $\pm$      2 &  -20.3 &  -25.2 &  -19.3 &   -24.6\\ 
30& gS0dS0o97    &  11.09 &    202 $\pm$      3 &  186.9 $\pm$    0.7 &    173 $\pm$      1 &&    205 $\pm$      3 &  187.9 $\pm$    0.7 &    174 $\pm$      1 &  -20.3 &  -25.2 &  -19.3 &   -24.6\\ 
31& gS0dS0o98    &  11.09 &    193 $\pm$      3 &  167.6 $\pm$    0.7 &    171 $\pm$      2 &&    195 $\pm$      3 &  168.1 $\pm$    0.7 &    171 $\pm$      1 &  -20.3 &  -25.2 &  -19.3 &   -24.6\\ 
32& gS0dS0o99    &  11.09 &    201 $\pm$      3 &  176.2 $\pm$    0.7 &    168 $\pm$      2 &&    203 $\pm$      3 &  177.0 $\pm$    0.7 &    167 $\pm$      2 &  -20.3 &  -25.2 &  -19.3 &   -24.6\\ 
33& gS0dSao10    &  11.09 &    208 $\pm$      3 &  206.0 $\pm$    0.7 &    171 $\pm$      2 &&    208 $\pm$      3 &  209.3 $\pm$    0.6 &    175 $\pm$      2 &  -20.3 &  -25.3 &  -19.3 &   -24.6\\ 
34& gS0dSao103$^{\dagger}$   &  11.09 &    216 $\pm$      6 &  205.6 $\pm$    0.7 &    168 $\pm$      2 &&    214 $\pm$      6 &  208.8 $\pm$    0.7 &    170 $\pm$      2 &  -20.3 &  -25.3 &  -19.3 &   -24.6\\ 
35& gS0dSao105$^{\dagger}$   &  11.09 &    225 $\pm$      8 &  194.8 $\pm$    0.7 &    170 $\pm$      2 &&    225 $\pm$      7 &  197.4 $\pm$    0.7 &    173 $\pm$      2 &  -20.3 &  -25.3 &  -19.3 &   -24.6\\ 
36& gS0dSao106$^{\dagger}$   &  11.09 &    224 $\pm$      8 &  207.2 $\pm$    0.7 &    166 $\pm$      2 &&    220 $\pm$      8 &  209.9 $\pm$    0.6 &    170 $\pm$      2 &  -20.3 &  -25.3 &  -19.4 &   -24.6\\ 
37& gS0dSbo106$^{\dagger}$   &  11.07 &    230 $\pm$      7 &  178.1 $\pm$    0.8 &    196 $\pm$      1 &&    245 $\pm$      5 &  178.0 $\pm$    0.7 &    198 $\pm$      1 &  -20.3 &  -25.2 &  -19.3 &   -24.6\\ 
38& gS0dSdo100   &  11.07 &    201 $\pm$      3 &  169.6 $\pm$    0.6 &    176 $\pm$      1 &&    197 $\pm$      3 &  174.7 $\pm$    0.6 &    175 $\pm$      1 &  -20.3 &  -25.2 &  -19.4 &   -24.6\\ 
39& gSagSao1$^{\dagger}$     &  11.37 &    242 $\pm$      9 &  221.3 $\pm$    2.4 &    170 $\pm$      3 &&    240 $\pm$      8 &  225.2 $\pm$    2.2 &    165 $\pm$      3 &  -21.3 &  -26.0 &  -19.9 &   -25.2\\ 
40& gSagSao5$^{\dagger}$     &  11.38 &    282 $\pm$     12 &  214.8 $\pm$    2.6 &    188 $\pm$      4 &&    281 $\pm$     12 &  224.9 $\pm$    0.8 &    184 $\pm$      4 &  -21.3 &  -26.1 &  -19.9 &   -25.2\\ 
41& gSagSao9$^{\dagger}$     &  11.38 &    238 $\pm$     17 &  192.7 $\pm$    2.1 &    206 $\pm$      3 &&    239 $\pm$     17 &  208.1 $\pm$    0.7 &    209 $\pm$      3 &  -21.2 &  -26.0 &  -19.9 &   -25.2\\ 
42& gSagSbo1$^{\dagger}$     &  11.26 &    263 $\pm$     11 &  201.7 $\pm$    2.2 &    157 $\pm$      3 &&    258 $\pm$     11 &  209.0 $\pm$    2.0 &    154 $\pm$      3 &  -21.1 &  -25.8 &  -19.8 &   -24.9\\ 
43& gSagSbo2     &  11.26 &      $\ldots$ &  201.8 $\pm$    2.3 &    144 $\pm$      3 &&      $\ldots$ &  209.9 $\pm$    0.7 &    132 $\pm$      3 &  -21.1 &  -25.8 &  -19.8 &   -24.9\\ 
44& gSagSbo21$^{\dagger}$    &  11.26 &    250 $\pm$      9 &  136.0 $\pm$    4.9 &    138 $\pm$      3 &&    250 $\pm$      8 &  132.7 $\pm$    4.9 &    122 $\pm$      3 &  -21.1 &  -25.7 &  -19.7 &   -24.9\\ 
45& gSagSbo22$^{\dagger}$    &  11.26 &    242 $\pm$      6 &  198.1 $\pm$    2.2 &    143 $\pm$      3 &&      $\ldots$ &  220.6 $\pm$    0.7 &    129 $\pm$      3 &  -21.1 &  -25.7 &  -19.7 &   -24.9\\ 
46& gSagSbo5$^{\dagger}$     &  11.26 &    232 $\pm$     15 &  194.1 $\pm$    2.2 &    173 $\pm$      4 &&    254 $\pm$     10 &  209.7 $\pm$    0.7 &    168 $\pm$      3 &  -21.1 &  -25.8 &  -19.7 &   -24.9\\ 
47& gSagSbo9$^{\dagger}$     &  11.26 &    204 $\pm$      5 &  189.2 $\pm$    0.6 &    183 $\pm$      3 &&    204 $\pm$      5 &  207.2 $\pm$    0.6 &    184 $\pm$      3 &  -21.1 &  -25.8 &  -19.7 &   -24.9\\ 
48& gSagSdo2$^{\dagger}$     &  11.27 &    222 $\pm$     11 &  187.4 $\pm$    2.4 &    177 $\pm$      3 &&    194 $\pm$     21 &  196.1 $\pm$    1.9 &    170 $\pm$      3 &  -21.4 &  -25.9 &  -20.0 &   -25.0\\ 
49& gSagSdo42$^{\dagger}$    &  11.27 &    248 $\pm$     10 &  179.0 $\pm$    2.3 &    182 $\pm$      3 &&    229 $\pm$     14 &  193.7 $\pm$    1.2 &    173 $\pm$      5 &  -21.2 &  -25.8 &  -19.9 &   -24.9\\ 
50& gSagSdo5$^{\dagger}$     &  11.27 &    249 $\pm$      9 &  178.9 $\pm$    1.8 &    179 $\pm$      3 &&    249 $\pm$     10 &  192.9 $\pm$    1.5 &    172 $\pm$      3 &  -21.3 &  -25.9 &  -20.0 &   -25.0\\ 
51& gSagSdo9$^{\dagger}$     &  11.28 &    190 $\pm$     11 &  187.8 $\pm$    2.1 &    180 $\pm$      3 &&    215 $\pm$      6 &  205.4 $\pm$    0.6 &    176 $\pm$      3 &  -21.3 &  -25.9 &  -19.9 &   -24.9\\ 
52& gSbgSbo22    &  11.10 &      $\ldots$ &  171.2 $\pm$    3.1 &     27 $\pm$      1 &&      $\ldots$ &  184.7 $\pm$    0.6 &     27 $\pm$      2 &  -20.8 &  -25.4 &  -19.6 &   -24.5\\ 
53& gSbgSbo72$^{\dagger}$    &  11.11 &    151 $\pm$      4 &  217.9 $\pm$    2.4 &     91 $\pm$      2 &&      $\ldots$ &  221.9 $\pm$    2.2 &     76 $\pm$      2 &  -20.8 &  -25.4 &  -19.5 &   -24.5\\ 
54& gSbgSbo9$^{\dagger}$     &  11.10 &    203 $\pm$      5 &  171.5 $\pm$    2.0 &    148 $\pm$      3 &&    187 $\pm$      7 &  177.5 $\pm$    0.5 &    142 $\pm$      3 &  -20.8 &  -25.4 &  -19.6 &   -24.5\\ 
55& gSbgSdo14$^{\dagger}$    &  11.12 &    148 $\pm$      4 &  141.4 $\pm$    2.3 &     85 $\pm$      4 &&    143 $\pm$      4 &  152.6 $\pm$    2.2 &     80 $\pm$      4 &  -21.0 &  -25.4 &  -19.7 &   -24.5\\ 
56& gSbgSdo17    &  11.12 &      $\ldots$ &  177.8 $\pm$    2.5 &     57 $\pm$      3 &&      $\ldots$ &  190.5 $\pm$    2.3 &     54 $\pm$      9 &  -21.0 &  -25.4 &  -19.7 &   -24.6\\ 
57& gSbgSdo18    &  11.13 &      $\ldots$ &  177.8 $\pm$    2.5 &     57 $\pm$      3 &&      $\ldots$ &  190.5 $\pm$    2.3 &     54 $\pm$      9 &  -21.0 &  -25.5 &  -19.7 &   -24.6\\ 
58& gSbgSdo19    &  11.12 &      $\ldots$ &  118.7 $\pm$    8.2 &     25 $\pm$      2 &&      $\ldots$ &  119.9 $\pm$    8.1 &     20 $\pm$      2 &  -21.0 &  -25.4 &  -19.8 &   -24.5\\ 
59& gSbgSdo41$^{\dagger}$    &  11.12 &    214 $\pm$      8 &  134.5 $\pm$    2.0 &    134 $\pm$      3 &&    211 $\pm$      7 &  145.9 $\pm$    1.9 &    121 $\pm$      3 &  -21.0 &  -25.4 &  -19.8 &   -24.6\\ 
60& gSbgSdo5$^{\dagger}$     &  11.12 &    245 $\pm$      9 &  168.4 $\pm$    2.9 &    163 $\pm$      4 &&    246 $\pm$      9 &  173.9 $\pm$    1.7 &    152 $\pm$      4 &  -21.1 &  -25.5 &  -19.7 &   -24.6\\ 
61& gSbgSdo69$^{\dagger}$    &  11.12 &    205 $\pm$      7 &  129.3 $\pm$    1.8 &    150 $\pm$      3 &&    203 $\pm$      7 &  135.8 $\pm$    1.6 &    140 $\pm$      3 &  -21.0 &  -25.4 &  -19.7 &   -24.5\\ 
62& gSbgSdo70$^{\dagger}$    &  11.12 &    163 $\pm$      4 &  154.1 $\pm$    1.8 &    145 $\pm$      2 &&    175 $\pm$      4 &  164.7 $\pm$    1.7 &    136 $\pm$      3 &  -21.0 &  -25.4 &  -19.8 &   -24.6\\ 
63& gSbgSdo9$^{\dagger}$     &  11.12 &    199 $\pm$     14 &  157.1 $\pm$    1.7 &    170 $\pm$      4 &&    225 $\pm$      9 &  169.9 $\pm$    1.7 &    164 $\pm$      4 &  -21.1 &  -25.5 &  -19.8 &   -24.6\\ 
64& gSdgSdo17    &  11.14 &      $\ldots$ &  154.6 $\pm$    4.3 &     28 $\pm$     10 &&      $\ldots$ &  165.4 $\pm$    3.1 &     24 $\pm$      4 &  -21.3 &  -25.6 &  -19.9 &   -24.6\\ 
65& gSdgSdo2     &  11.14 &      $\ldots$ &  176.1 $\pm$    4.1 &    137 $\pm$      2 &&      $\ldots$ &  181.6 $\pm$    3.3 &    118 $\pm$      2 &  -21.4 &  -25.7 &  -20.0 &   -24.6\\ 
66& gSdgSdo42$^{\dagger}$    &  11.14 &    184 $\pm$     12 &  152.3 $\pm$    2.3 &    144 $\pm$      2 &&    198 $\pm$      7 &  166.0 $\pm$    2.1 &    123 $\pm$      2 &  -21.3 &  -25.6 &  -20.0 &   -24.6\\ 
67& gSdgSdo5$^{\dagger}$     &  11.14 &    150 $\pm$     10 &  150.5 $\pm$    2.3 &    154 $\pm$      2 &&    184 $\pm$      5 &  163.1 $\pm$    2.2 &    134 $\pm$      2 &  -21.4 &  -25.6 &  -20.0 &   -24.7\\ 
68& gSdgSdo69$^{\dagger}$    &  11.14 &    225 $\pm$      9 &  120.2 $\pm$    2.0 &    138 $\pm$      2 &&    223 $\pm$      9 &  126.9 $\pm$    1.7 &    127 $\pm$      3 &  -21.2 &  -25.5 &  -19.9 &   -24.6\\ 
69& gSdgSdo71    &  11.14 &      $\ldots$ &  139.9 $\pm$    2.3 &     98 $\pm$      5 &&      $\ldots$ &  139.9 $\pm$    1.5 &     95 $\pm$      4 &  -21.3 &  -25.6 &  -20.0 &   -24.6\\ 
70& gSdgSdo74    &  11.15 &      $\ldots$ &  192.7 $\pm$    0.4 &     42 $\pm$      5 &&      $\ldots$ &  198.4 $\pm$    0.4 &     39 $\pm$      8 &  -21.3 &  -25.6 &  -19.9 &   -24.6\\ 
71& gSdgSdo9$^{\dagger}$     &  11.15 &    228 $\pm$     10 &  151.4 $\pm$    0.4 &    160 $\pm$      2 &&    229 $\pm$      9 &  162.9 $\pm$    0.4 &    138 $\pm$      2 &  -21.4 &  -25.6 &  -20.0 &   -24.6\\ 
\hline
\end{longtable}
\end{center}
}
\vspace{-0.8cm}
\tablefoot{\emph{Columns}: (1) Number ID.  (2) Code g[\textit{type$_{1}$}]g[\textit{type$_{2}$}]o[\textit{orbit}] of the experiment generating an S0 remnant as explained in \S\ref{secc:magnitudes}. (3) Logarithm of the stellar mass of the remnant at the last moment of the simulation. (4,5,6) Circular velocity, central velocity dispersion, and maximum rotational velocity of the data weighted by the luminosity in the V band of the stellar content of the remnants, respectively. (7,8,9) Circular velocity, central velocity dispersion, and maximum rotational velocity for the values obtained weighting the data by the stellar mass of the particles in the remnant. (10) Total absolute magnitude in the Johnson-Cousin $B$ band at $z=0.8$. (11) Total absolute magnitude in the $K$ band at $z=0.8$. (12) Total absolute magnitude in the Johnson-Cousin $B$ band at $z=0$. (13) Total absolute magnitude in the $K$ band at $z=0$. The magnitudes are in the Vega system and have been corrected for dust extinction (see \S\ref{secc:corrections}).$^{\dagger}$ The circular velocity of these S0-like remnants can only be considered as an upper limit (see \S\ref{secc:calculovcirc} for the explanation).
}
\end{landscape}
\end{longtab}
\twocolumn

\end{appendix}

\end{document}